\documentclass[10pt]{amsart}
\usepackage{graphicx}
\usepackage{amssymb}
\title{Bayesian inference of natural selection from allele frequency time series}
\author{Joshua G. Schraiber}
\author{Steven N. Evans}
\author{Montgomery Slatkin}
\thanks{JGS supported by NSF grant DBI-1402120, SNE supported in part by NSF grant DMS-0907630, NSF grant DMS-1512933, and NIH grant
1R01GM109454-01, MS supported by NIH grant  R01-GM40282}
\date{Started on December 10, 2013. Compiled on \today}                                           

\begin{document}

\begin{abstract}
The advent of accessible ancient DNA technology now allows the direct ascertainment of allele frequencies in ancestral populations, thereby enabling the use of allele frequency time series to detect and estimate natural selection. Such direct observations of allele frequency dynamics are expected to be more powerful than inferences made using patterns of linked neutral variation obtained from modern individuals. We developed a Bayesian method to make use of allele frequency time series data and infer the parameters of general diploid selection, along with allele age, in non-equilibrium populations. We introduce a novel path augmentation approach, in which we use Markov chain Monte Carlo to integrate over the space of allele frequency trajectories consistent with the observed data. Using simulations, we show that this approach has good power to estimate selection coefficients and allele age. Moreover, when applying our approach to data on horse coat color, we find that ignoring a relevant demographic history can significantly bias the results of inference. Our approach is made available in a C++ software package. 
\end{abstract}

\maketitle

\section{Introduction}
The ability to obtain high-quality genetic data from ancient samples is revolutionizing the way that we understand the evolutionary history of populations. One of the most powerful applications of ancient DNA (aDNA) is to study the action of natural selection. While methods making use of only modern DNA sequences have successfully identified loci evolving subject to natural selection \cite{nielsen2005genomic,voight2006map,pickrell2009signals}, they are inherently limited because they look indirectly for selection, finding its signature in nearby neutral variation. In contrast, by sequencing ancient individuals, it is possible to directly track the change in allele frequency that is characteristic of the action of natural selection. This approach has been exploited recently using whole genome data to identify candidate loci under selection in European humans \cite{mathieson2015genome}.

To infer the action of natural selection rigorously, several methods have been developed to explicitly fit a population genetic model to a time series of allele frequencies obtained via aDNA. Initially, \cite{bollback2008estimation} extended an approach devised by \cite{williamson1999using} to estimate the population-scaled selection coefficient, $\alpha = 2N_e s$, along with the effective size, $N_e$. To incorporate natural selection, \cite{bollback2008estimation} used the continuous diffusion approximation to the discrete Wright-Fisher model. This required them to use numerical techniques to solve the partial differential equation (PDE) associated with transition densities of the diffusion approximation to calculate the probabilities of the population allele frequencies at each time point. \cite{ludwig2009coat} obtained an aDNA time series from 6 coat-color-related loci in horses and applied the method of \cite{bollback2008estimation} to find that 2 of them, ASIP and MC1R, showed evidence of strong positive selection.

Recently, a number of methods have been proposed to extend the generality of the \cite{bollback2008estimation} framework. To define the hidden Markov model they use, \cite{bollback2008estimation} were required to posit a prior distribution on the allele frequency at the first time point. They chose to use a uniform prior on the initial frequency; however, in truth the initial allele frequency is dictated by the fact that the allele at some point arose as a new mutation. Using this information, \cite{malaspinas2012estimating} developed a method that also infers allele age. They also extended the selection model of \cite{bollback2008estimation} to include fully recessive fitness effects. A more general selective model was implemented by \cite{steinrucken2014novel}, who model general diploid selection, and hence they are able to fit data where selection acts in an over- or under-dominant fashion; however, \cite{steinrucken2014novel} assumed a model with recurrent mutation and hence could not estimate allele age. The work of \cite{mathieson2013estimating} is designed for inference of metapopulations over short time scales and so it is computationally feasible for them to use a discrete time, finite population Wright-Fisher model. Finally, the approach of \cite{feder2014identifying} is ideally suited to experimental evolution studies because they work in a strong selection, weak drift limit that is common in evolving microbial populations. 

One key way that these methods differ from each other is in how they compute the probability of the underlying allele frequency changes. For instance, \cite{malaspinas2012estimating} approximated the diffusion with a birth-death type Markov chain, while \cite{steinrucken2014novel} approximate the likelihood analytically using a spectral representation of the diffusion discovered by \cite{song2012simple}. These different computational strategies are necessary because of the inherent difficulty in solving the Wright-Fisher partial differential equation. A different approach, used by \cite{mathieson2013estimating} in the context of a densely-sampled discrete Wright-Fisher model, is to instead compute the probability of the entire allele frequency trajectory in between sampling times.

In this work, we develop a novel approach for inference of general diploid selection and allele age from allele frequency time series obtained from aDNA. The key innovation of our approach is that we impute the allele frequency trajectory between sampled points when they are sparsely-sampled. Moreover, by working with a diffusion approximation, we are able to easily incorporate general diploid selection and changing population size. This approach to inferring parameters from a sparsely-sampled diffusion is known as high-frequency path augmentation, and has been successfully applied in a number of contexts \cite{roberts2001inference, golightly2005bayesian, golightly2008bayesian, sorensen2009parametric, fuchs2013inference}. The diffusion approximation to the Wright-Fisher model, however, has several features that are atypical in the context of high-frequency path augmentation, including a time-dependent diffusion coefficient and a bounded state-space. We then apply this new method to several datasets and find that we have power to estimate parameters of interest from real data.

\section{Model and Methods}
\subsection{Generative model}
We assume a randomly mating diploid population that is size $N(t)$ at time $t$, where $t$ is measured in units of $2N_0$ generations for some arbitrary, constant $N_0$. At the locus of interest, the ancestral allele, $A_0$, was fixed until some time $t_0$ when the derived allele, $A_1$, arose with diploid fitnesses as given in Table 1.
\begin{table}[h]
\centering
\begin{tabular} { c | c | c | c }
Genotype & $A_1A_1$ & $A_1A_0$ & $A_0A_0$ \\ \hline
Fitness & $1 + s_2$ & $1+ s_1$ & $1$
\end{tabular}
\caption{Fitness scheme assumed in the text.}
\end{table}

Given that an allele is segregating at a population frequency $0 < x_* < 1$ at some time $t_* > t_0$, 
the trajectory of population frequencies of $A_1$ at times $t  \ge t_*$, $(X_t)_{t \ge t_*}$, 
is modeled by the usual diffusion approximation to the Wright-Fisher model 
(and many other models such as the Moran model), which we will henceforth call the Wright-Fisher diffusion. 
While many treatments of the Wright-Fisher diffusion define it in terms of the 
partial differential equation that characterizes
its transition densities
(e.g. \cite{ewens2004mathematical}), 
we instead describe it as the solution of a stochastic differential equation (SDE). 
Specifically, $(X_t)_{t \ge t_*}$ satisfies the SDE
\begin{equation}
\begin{split}
dX_t & = X_t(1-X_t)(\alpha_1(2X_t-1)-\alpha_2 X_t) \, dt + \sqrt{\frac{X_t(1-X_t)}{\rho(t)}} \, dB_t \\
X_{t_*} & = x_*,\\
\end{split}
\label{WF_SDE}
\end{equation}
where $B$ is a standard Brownian motion, $\alpha_1 = 2N_0 s_1$, $\alpha_2 = 2N_0 s_2$, and  $\rho(t) = N(t)/N_0$. 
If $X_{t_{**}} = 0$ (resp. $X_{t_{**}} = 1$) at some time $t_{**} > t_*$,
then $X_t = 0$ (resp. $X_t = 1$) for all $t \ge t_{**}$.

In order to make this description of the dynamics of the population allele frequency trajectory
$(X_t)_{t \ge t_0}$ complete, we need to specify an initial condition at time $t_0$.  In a
finite population Wright-Fisher model we would take the allele $A_1$ to have frequency
$\frac{1}{2N(t_0)}$ at the time $t_0$ when it first arose in a single chromosome.  
This frequency converges to $0$ when we pass to the
diffusion limit, but we cannot start the Wright-Fisher diffusion at $0$ at time $t_0$ because
the diffusion started at $0$ remains at $0$.  Instead, we take the value of $X_{t_0}$ to be some
small, but arbitrary, frequency $x_0$.  This arbitrariness in the choice of $x_0$ may seem
unsatisfactory, but we will see that the resulting posterior distribution for the
parameters $\alpha_1, \alpha_2, t_0$ converges
as $x_0 \downarrow 0$ to a  limit which can be thought of as the posterior
corresponding to a certain improper prior distribution, and so, in the end, 
there is actually no need to specify $x_0$.

Finally, we model the data assuming that at known times $t_1, t_2, \ldots, t_k$ samples of known sizes
$n_1, n_2, \ldots, n_k$ chromosomes are taken and 
$c_1, c_2, \ldots, c_k$  copies of the derived allele are found at the successive time points (Figure 1). 
Note that it is possible that some of the sampling times are more ancient than $t_0$, the age of the allele. 

\begin{figure}[h] 
   \includegraphics[width=.9\textwidth]{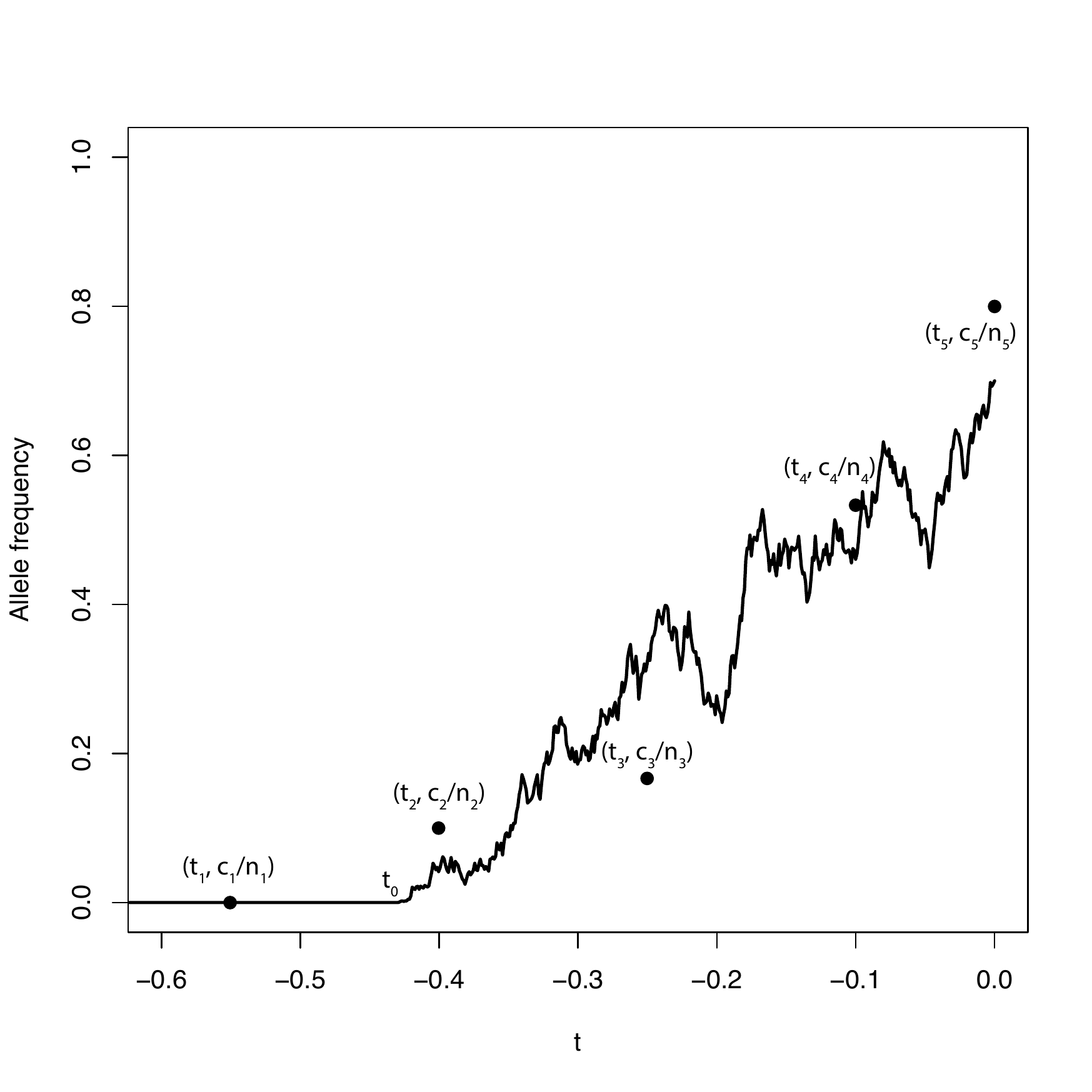} 
   \caption{Taking samples from an allele frequency trajectory. An allele frequency trajectory is simulated from the Wright-Fisher diffusion (solid line). At each time, $t_i$, a sample of size $n_i$ chromosomes is taken and $c_i$ copies of the derived allele are observed. Each point corresponds to the observed allele frequency of sample $i$. Note that $t_1$ is more ancient than the allele age, $t_0$. }
\end{figure}

\subsection{Bayesian path augmentation}
\label{SS:path_augmentation}
We are interested in devising a Bayesian method to obtain 
the posterior distribution on the parameters, $\alpha_1$, 
$\alpha_2$, and $t_0$ given the sampled allele frequencies and sample times 
-- data which we denote collectively as $D$.
Because we are dealing with objects that don't necessarily have distributions which have densities with respect
to canonical reference measures, it will be convenient in the beginning
to treat priors and posteriors as probability measures
rather than as density functions. 
For example, the posterior is the probability measure
\begin{equation}
\label{posterior_naive}
P(d \alpha_1, d \alpha_2, d t_0 \, | \,D) 
= 
\frac{P(d D \, | \,\alpha_1, \alpha_2, t_0) \, \pi(d \alpha_1, d \alpha_2, d t_0)}{P(d D)},
\end{equation} 
where $\pi$ is a joint prior on the model parameters. However, computing 
the likelihood $P(d D \, | \,\alpha_1, \alpha_2, t_0)$ is computationally challenging because, implicitly,
\[
P(d D \, | \,\alpha_1, \alpha_2, t_0) = \int P(d D \, | \,X) \, P(dX \, | \,\alpha_1, \alpha_2, t_0),
\]
where the integral is over the (unobserved, infinite-dimensional) allele frequency path $X = (X_t)_{t \ge t_0}$,
$P(\cdot \, | \,\alpha_1, \alpha_2, t_0)$ is the distribution of a Wright-Fisher diffusion with
selection parameters $\alpha_1, \alpha_2$ started at time $t_0$ at the small but arbitrary frequency $x_0$,
and 
\[
P(d D \, | \,X) = \prod_{i=1}^k \binom{n_i}{c_i}X_{t_i}^{c_i}(1-X_{t_i})^{n_i-c_i}
\]
because we assume that sampled allele frequencies at the times $t_1, \ldots, t_k$
are independent binomial draws governed by underlying population allele frequencies at the these times.
Integrating over the infinite-dimensional path $(X_t)_{t  \ge t_0}$ involves either 
solving partial differential equations numerically
or using Monte Carlo methods to find the joint distribution of population allele frequency path at the
times $t_1, \ldots, t_k$.

 To address this computational difficulty, we introduce a path augmentation method
that treats the underlying 
allele frequency path $(X_t)_{t \ge t_0}$ as an additional parameter.  
Observe that the posterior may be expanded out to
\[
P(d\alpha_1, d\alpha_2, d t_0 \, | \,D) 
=
\frac{\int P(dD \, | \,X') \, P(dX' \, | \,\alpha_1, \alpha_2, t_0) \pi( d\alpha_1, d\alpha_2, dt_0)}
{\int P(dD \, | \,X') \, P(dX' \, | \,\alpha_1', \alpha_2', t_0') \pi(d \alpha_1', d \alpha_2', d t_0')},
\]
where we use primes to designate dummy variables over which we integrate.
Thinking of the path $(X_t)_{t  \ge t_0}$ as another parameter and taking the prior distribution
for the augmented family of parameters to be
\[
P(dX \, | \,\alpha_1, \alpha_2, t_0) \pi( d\alpha_1, d\alpha_2, dt_0),
\]
the posterior for the augmented family of parameters is
\begin{equation}
\label{posterior}
P(d\alpha_1, d\alpha_2, dt_0; dX \, | \,D) 
= 
\frac
{P(dD \, | \,X) P(d X \, | \,\alpha_1, \alpha_2, t_0) \pi(d\alpha_1, d\alpha_2, dt_0)}
{\int P(dD \, | \,X') P(dX' \, | \,\alpha_1', \alpha_2', t_0') \pi(d \alpha_1', d\alpha_2', dt_0')}.
\end{equation}

We thus see that treating the allele frequency path as a parameter is consistent with
the initial ``naive'' Bayesian approach in that if we integrate the path variable out of the
posterior \eqref{posterior} for the augmented family of parameters, then we recover the
posterior \eqref{posterior_naive} for the original family of parameters.  
In practice, this means that marginalizing out the path variable
from a Monte Carlo approximation of the augmented posterior gives a Monte Carlo
approximation of the original posterior.

Implicit in our set-up is the initial frequency $x_0$ at time $t_0$. Under the probability
measure governing the Wright-Fisher diffusion, any process started from $x_0 = 0$ will
stay there forever. Thus, we would be forced to make an arbitrary choice of some $x_0 > 0$ 
as the initial frequency of our allele. However, we argue in the Appendix that in the limit as
$x_0 \downarrow 0$, we can achieve an improper prior distribution on the space of allele
frequency trajectories. We stress that our inference using such an improper prior 
is not one that arises directly from a generative probability model
for the allele frequency path. However, it does arise as a limit as the initial allele frequency
$x_0$ goes to zero of inferential procedures based on generative probability models
and the limiting posterior distributions are probability distributions.  
Therefore, the parameters $\alpha_1, \alpha_2, t_0$ retain their meaning,
our conclusions can be thought of approximations to 
those that we would arrive at for all sufficiently small values of $x_0$, 
and we are spared the necessity of making an arbitrary choice of $x_0$.

\subsection{Path likelihoods}
\label{SS:path_likelihoods}
Most instances of Bayesian inference in population genetics have hitherto involved
finite-dimensional parameters. We recall that if a finite-dimensional parameter has 
a diffuse prior distribution (that is,
a distribution where an individual specification of values of the parameter has zero prior probability), 
then one replaces the prior probabilities of parameter values that would be appear
when if we had a discrete prior distribution by evaluations of densities with respect 
to an underlying reference measure -- usually Lebesgue measure in an appropriate dimension 
-- and the Bayesian formalism then proceeds 
in much the same way as it does in the discrete case with, 
for example, ratios of probabilities replaced by ratios of densities.
We thus require a reference measure on the infinite-dimensional space of paths
that will play a role analogous to that of Lebesgue measure in the finite-dimensional case.

To see what is involved, suppose we have a diffusion process $(Z_t)_{t \ge t_0}$ 
that satisfies the SDE
\begin{equation}
\begin{split}
dZ_t & = a(Z_t, t) \, dt + dB_t \\
Z_{t_0} & = z_0, \\
\end{split}
\label{generic_SDE}
\end{equation}
where $B$ is a standard Brownian motion (the Wright-Fisher diffusion is not of this form but, as we
shall soon see, it can be be reduced to it after suitable transformations of time and space).
Let $\mathbb{P}$ be the distribution of $(Z_t)_{t \ge t_0}$ -- this is a probability
distribution on the space of continuous paths that start from position $z_0$ at time $t_0$. 
While the probability assigned by $\mathbb{P}$ to any particular path is zero, 
we can, under appropriate conditions, make sense of the probability of
a path under $\mathbb{P}$ \emph{relative} to its probability under the distribution of Brownian motion. 
If we denote by $\mathbb{W}$ the distribution of Brownian motion starting from position
$z_0$ at time $t_0$, then Girsanov's theorem \cite{girsanov1960transforming} 
gives the density of the path segment $(Z_s)_{t_0 \leq s \leq t}$ under $\mathbb{P}$ relative to $\mathbb{W}$ as
\begin{equation}
\frac{d\mathbb{P}}{d\mathbb{W}}((Z_s)_{t_0 \leq s \leq t}) 
= \exp\left\{ \int_{t_0}^t a(Z_s, s) \, dZ_s 
- \frac{1}{2}\int_{t_0}^t a^2(Z_s, s) \, ds \right\},
\label{girsanov}
\end{equation}
where the first integral in the exponentiand is an It\^o integral.  In order for \eqref{girsanov} to hold,
the integral $\int_{t_0}^t a^2(Z_s, s) \, ds$ must be finite, in which case the It\^o integral 
$\int_{t_0}^t a(Z_s, s) \, dZ_s$ is also well-defined and finite.

However, the Wright-Fisher SDE \eqref{WF_SDE} is not of the form \eqref{generic_SDE}. 
In particular, the factor multiplying the infinitesimal Brownian increment $dB_t$ 
(the so-called diffusion coefficient) depends on both space and time. 
To deal with this issue, we first apply a well-known time transformation 
and consider the process $(\tilde X_\tau)_{\tau \ge 0}$
given by $\tilde X_\tau = X_{f^{-1}(\tau)}$, where
\begin{equation}
\label{f_def}
f(t) = \int_{t_0}^t \frac{1}{\rho(s)} \, ds, \quad t \ge t_0.
\end{equation}

It is not hard to see that $(\tilde X_\tau)_{\tau \ge 0}$ satisfies the
following SDE with a time-independent diffusion coefficient,
\[
\begin{split}
d \tilde X_\tau 
& = \rho(f^{-1}(\tau)) \tilde X_\tau(1 - \tilde X_\tau)(\alpha_1(2 \tilde X_\tau - 1)-\alpha_2 \tilde X_\tau) \, d\tau 
+ \sqrt{\tilde X_\tau(1- \tilde X_\tau)} \, d \tilde B_\tau \\
\tilde X_0 & = x_0, \\
\end{split}
\]
where $\tilde B$ is a standard Brownian motion.
Next, we employ an angular space transformation first suggested by \cite{fisher1922dominance}, 
$Y_\tau = \arccos(1-2 \tilde X_\tau)$. 
Applying It\^o's lemma \cite{ito1944stochastic} shows that $(Y_\tau)_{\tau \ge 0}$ is a diffusion 
that satisfies the SDE
\begin{equation}
\begin{split}
dY_\tau & = \frac{1}{4}\left(\rho(f^{-1}(\tau))\sin(Y_\tau)(\alpha_2+(2\alpha_1-\alpha_2)\cos(Y_\tau)) 
- 2\cot(Y_\tau)\right) \, d\tau + dW_\tau \\
Y_0 & = y_0 = \arccos(1-2 x_0), \\
\end{split}
\label{WF_SDE_FISHER}
\end{equation}
where $W$ is a standard Brownian motion.  If the process $X$ hits either of the boundary points $0,1$,
then it stays there, and the same is true of the time and space transformed process $Y$ 
for its boundary points $0, \pi$.

The restriction of the distribution of the time and space transformed process
$Y$ to some set of paths that don't hit the boundary
is absolutely continuous with respect to the distribution of standard Brownian motion restricted to the
same set; that is, the distribution of $Y$ restricted to such a set of paths
has a density with respect to the distribution of Brownian motion restricted to the same set.
However, the infinitesimal mean in \eqref{WF_SDE_FISHER} 
(that is, the term multiplying $d \tau$) becomes singular as $Y_\tau$ approaches the boundary points 
$0$ and $\pi$, corresponding to the boundary points $0$ and $1$ for allele frequencies. 
These singularities prevent the process $Y$ from
re-entering the interior of its state space and ensure that a Wright-Fisher path will be absorbed when the allele is either fixed or lost. A consequence is that the density of the distribution of $Y$
relative to that of a Brownian motion blows up as the path approaches the boundary.  We are modeling the
appearance of a new mutation in terms of a Wright-Fisher diffusion starting at some small
initial frequency $x_0$ at time $t_0$ and we want to perform our parameter inference in such
a way that we get meaningful answers as $x_0 \downarrow 0$.  This suggests that rather than
working with the distribution $\mathbb{W}$ of Brownian motion as a reference measure
it may be more appropriate to work with a tractable diffusion process that
exhibits similar behavior near the boundary point $0$.

To start making this idea of matching singularities more precise,
consider a diffusion process $(\bar Z_t)_{t \ge t_0}$ that satisfies the SDE
\begin{equation}
\begin{split}
d \bar Z_t & = b(\bar Z_t, t) \, dt + d \bar B_t \\
\bar Z_0 & = z_0, \\
\end{split}
\label{generic_SDE_b}
\end{equation}
where $\bar B$ is a standard Brownian motion.
Write $\mathbb{Q}$ for the distribution of the diffusion process $(\bar Z_t)_{t \ge t_0}$
and recall that $\mathbb{P}$ is the distribution of a solution of \eqref{generic_SDE}.
If $(Z_s)_{t_0 \leq s \leq t}$ is a segment of path such that both $\int_{t_0}^t a^2(Z_s, s) \, ds < \infty$
and $\int_{t_0}^t b^2(Z_s, s) \, ds < \infty$, then
\begin{align}
\frac{d\mathbb{P}}{d\mathbb{Q}}((Z_s)_{t_0\leq s \leq t}) 
&= \frac{d\mathbb{P}}{d\mathbb{W}}((X_s)_{t_0\leq s \leq t})
\Big/ 
\frac{d\mathbb{Q}}{d\mathbb{W}}((Z_s)_{t_0\leq s \leq t}) \nonumber \\
&= \exp\left\{\int_{t_0}^t \left(a(Z_s, s) - b(Z_s, s)\right) \, dZ_s 
- \frac{1}{2}\int_{t_0}^t \left(a^2(Z_s, s) - b^2(Z_s, s) \right) \, ds \right\}.
\label{girsanovOther}
\end{align}
Note that the right-hand side will stay bounded if one considers a sequence of paths, indexed by $\eta$,
$(Z_s^\eta)_{t_0\leq s \leq t}$, with 
$\int_{t_0}^t a^2(Z_s^\eta, s) \, ds < \infty$
and $\int_{t_0}^t b^2(Z_s^\eta, s) \, ds < \infty$, provided that
$\int_{t_0}^t (a^2(Z_s^\eta, s) - b^2(Z_s^\eta, s)) \, ds$
stays bounded.  These manipulations with densities may seem somewhat heuristic,
but they can be made rigorous and, moreover, the form of $\frac{d \mathbb{P}}{d \mathbb{Q}}$
follows from an extension of Girsanov's theorem that gives the density of $\mathbb{P}$
with respect to $\mathbb{Q}$ directly without using the densities with respect to $\mathbb{W}$ 
as intermediaries
(see, for example, \cite[Theorem~18.10]{MR1876169}).

We wish to apply this observation to the time and space transformed Wright-Fisher diffusion
of \eqref{WF_SDE_FISHER}.  Because
\[
-\frac{1}{2} \cot(y) 
+ 
\frac{1}{4}\rho(f^{-1}(t)) \sin(y)
\left((2 \alpha_1 - \alpha_2) \cos(y) + \alpha_2\right)
= -\frac{1}{2y} + O(y) 
\]
when $y$ is small, an appropriate reference process should have infinitesimal mean $b(y,t) \approx -1/(2y)$ as $y \downarrow 0$. Following suggestions by \cite{schraiber2013analysis} and \cite{jenkins2013exact}, we compute path densities relative to the distribution $\mathbb{Q}$ of the Bessel(0) process, a process
which is the solution of the SDE 
\begin{equation}
\label{Bessel_0_SDE}
\begin{split}
d \bar Y_t & = -\frac{1}{2 \bar Y_t} \, dt + d \bar B_t, \\
\bar Y_0 & = y_0 = \arccos(1-2 x_0). \\
\end{split}
\end{equation}
For the moment, write $\mathbb{P}^{y_0}$ and $\mathbb{Q}^{y_0}$ for the respective distributions
of the solutions of \eqref{WF_SDE_FISHER} and \eqref{Bessel_0_SDE} to emphasize the dependence on $y_0$
(equivalently, on the initial allele frequency $x_0$).  
There are $\sigma$-finite measures $\mathbb{P}^0$ and $\mathbb{Q}^0$ with infinite total mass such that
for each $\epsilon > 0$ 
\[
\lim_{y_0 \downarrow 0} \mathbb{P}^{y_0}((Y_t)_{t \ge \epsilon} \in \cdot \, | \,Y_\epsilon > 0)
=
\mathbb{P}^0((Y_t)_{t \ge \epsilon} \in \cdot) \Big / \mathbb{P}^0(Y_\epsilon > 0)
\]
and
\[
\lim_{y_0 \downarrow 0} \mathbb{Q}^{y_0}((\bar Y_t)_{t \ge \epsilon} \in \cdot \, | \,\bar Y_\epsilon > 0)
=
\mathbb{Q}^0((\bar Y_t)_{t \ge \epsilon} \in \cdot) \Big / \mathbb{Q}^0(\bar Y_\epsilon > 0),
\]
where the numerators and denominators in the last two equations are all finite.
Moreover, $\mathbb{P}^0$ has a density with respect to $\mathbb{Q}^0$ that arises by naively taking
limits as $y_0 \downarrow 0$
in the functional form of the density of $\mathbb{P}^{y_0}$ with respect to $\mathbb{Q}^{y_0}$
(we say ``naively'' because $\mathbb{P}^{y_0}$ and $\mathbb{Q}^{y_0}$ assign all of their mass
to paths that start at position $y_0 = \arccos(1-2 x_0)$ at time $0$, whereas $\mathbb{P}^0$ and $\mathbb{Q}^0$
assign all of their mass to paths that start at position $0$ at time $0$, and so the set of paths
at which it is relevant to compute the density changes as $y_0 \downarrow 0$).  As we have already remarked,
the limit of our Bayesian inferential procedure may be thought of as Bayesian inference with an improper prior, 
but we stress that the resulting posterior is proper. 

The notion of the infinite measure $\mathbb{Q}^0$ may seem somewhat forbidding, 
but this measure is characterized by the following simple properties:
\[
\mathbb{Q}^0(\bar Y_\epsilon \in dy) 
= 
\frac{y^2}{\epsilon^2} \exp\left\{-\frac{y^2}{2 \epsilon}\right\} \, dy,
\quad y > 0,
\]
so that $\mathbb{Q}^0(\bar Y_\epsilon > 0) = \sqrt{\frac{\pi}{2}} \frac{1}{\sqrt{\epsilon}}$,
and conditional on the event $\{\bar Y_\epsilon = y\}$ the evolution
of $(\bar Y_t)_{t \ge \epsilon}$ is exactly that of the Bessel(0)
process started at position $y$ at time $\epsilon$.  Moreover, conditional
on the event $\{\bar Y_s = a, \, \bar Y_u = b\}$ for $0 \le s < u$ and $a,b > 0$, the evolution
of the ``bridge'' $(\bar Y_u)_{s \le t \le u}$ is the same as that
of the corresponding bridge for a Bessel(4) process; a Bessel(4) process satisfies the SDE
\[
d \hat Y_t = \frac{3}{2 \hat Y_t} \, dt + d \hat B_t.
\]
Very importantly for the sake of simulations, the Bessel(4) process is just the radial
part of a $4$-dimensional standard Brownian motion -- in particular, this process
started at $0$ leaves immediately and never returns.  Also, the Bessel(0) process
arises naturally because our space transformation 
$x \mapsto \arccos(1-2 x) = \int_0^x \frac{1}{\sqrt{w(1-w)}} \, dw$ is approximately 
$x \mapsto \int_0^x \frac{1}{\sqrt{w}} \, dw = 2 \sqrt x$ when $x > 0$ is small and a multiple of the
square of Bessel(0) process, sometimes called Feller's continuous state branching
processes, arises naturally as an approximation to the Wright-Fisher diffusion for
low frequencies \cite{haldane1927mathematical, feller1951diffusion}.

\subsection{The joint likelihood of the data and the path}
To write down down the full likelihood of the observations and the path, 
we make the assumption that 
the population size function $\rho(t)$ is continuously differentiable except at a finite set of times 
$d_1 < d_2 < \ldots < d_M$. 
Further, we require that  that $\rho(d_i^+) = \lim_{t \downarrow d_i}\rho(t)$ exists and is equal to 
$\rho(d_i)$ while $\rho(d_i^-) = \lim_{t\uparrow d_i}\rho(t)$ also exists 
(though it may not necessarily equal $\rho(d_i)$). 

Using the notation of Subsection~\ref{SS:path_augmentation},
write 
\[
L(D, (Y_t)_{t \ge 0} \, | \, \alpha_1, \alpha_2, t_0) 
= \mathbb{P}(D \, | \, (Y_t)_{t \ge 0}, t_0) \Phi^0((Y_t)_{t \ge 0}; \alpha_1, \alpha_2, t_0)
\]
for the joint likelihood of the data and the time and space
transformed allele frequency path $(Y_t)_{t \ge 0}$ given the
parameters $\alpha_1, \alpha_2, t_0$.
In the Appendix, we show that 
\begin{align}
\begin{split}
& L(D, (Y_s)_{0 \leq s \leq t_k} \, | \,\alpha_1, \alpha_2, t_0)  \\
& \quad =  \exp\left. \bigg\{A(Y_{f(t_k)},t_k^-) +A(Y_{f(d_m)}, d_m^-) - (A(Y_{f(d_K)},d_K) + A(Y_{f(t_0)},t_0)) \right.  \\
& \qquad \left.  + \sum_{i=m}^{K} \left[A(Y_{f(d_{i+1})},d_{i+1}^-) - A(Y_{f(d_i)},d_i) \right] \right.  \\
& \qquad \left.  - \int_{t_0}^{t_k} B(Y_{f(s)},s) ds - \frac{1}{2}\int_{t_0}^{t_k} C(Y_{f(s)},s)ds - \frac{1}{2}\int_{t_0}^{t_k} D(Y_{f(s)}, s) ds \right\} \\
&  \qquad \times \prod_{i=1}^k \binom{n_i}{c_i}\left(\frac{1-\cos(Y_{f(t_i)})}{2} \right)^{c_i}\left( \frac{1+ \cos(Y_{f(t_i)})}{2}\right)^{n_i-c_i}, 
\end{split}
\label{wholelikelihood}
\end{align}
where $f$ is as in \eqref{f_def}, $m = \min \{ i : d_i > t_0\}$ and $K = \max\{ i : d_i > t_k\}$, and 
\begin{align*}
A(y,t) &= \frac{\log(y)}{2} - \frac{1}{8}\left(\rho(t)\cos(y)(2\alpha_2+(2\alpha_1-\alpha_2)\cos(y)) + 4\log(\sin(y))\right) \\
B(y,t) &= -\frac{1}{8}\frac{d\rho}{dt}(t)\cos(y)(2\alpha_2+(2\alpha_1-\alpha_2)\cos(y)) \\
C(y,t) &= \frac{1}{2}\left(\alpha_1\cos(y) + \frac{\csc(y)^2}{\rho(t)}\right) - \frac{1}{2y^2\rho(t)} \\
D(y,t) &= \frac{1}{16\rho(t)}\left(\rho(t)\sin(y)(\alpha_2+(2\alpha_1-\alpha_2)\cos(y))-2\cot(y)\right)^2-\frac{1}{4y^2\rho(t)}.
\end{align*}
While this expression may appear complicated, it has the important feature that, unlike the form of the
likelihood that would arise by simply applying Girsanov's theorem, it only involves 
Lebesgue (indeed Riemann) integrals
and not It\^o integrals, which, as we recall in the Appendix, are known 
from the literature to be potentially difficult to compute numerically.

\subsection{Metropolis-Hastings algorithm}
We now describe a Markov chain Monte Carlo method for Bayesian inference of the parameters 
$\alpha_1$, $\alpha_2$ and $t_0$, along with the allele frequency path $(X_t)_{t \ge t_0}$
(equivalently, the transformed path $(Y_t)_{t\ge 0}$). 
While updates to the selection parameters $\alpha_1$ and $\alpha_2$ do not require updating the path, 
updating the time $t_0$ at which the derived allele arose
requires proposing updates to the segment of path from $t_0$ up to the
time of the first sample with a non-zero number of derived alleles.
Additionally, we require proposals to update small sections of the path without updating 
any parameters and proposals to update the allele frequency at the most recent sample time.

\begin{figure}[h] 
   \includegraphics[width=1.1\textwidth]{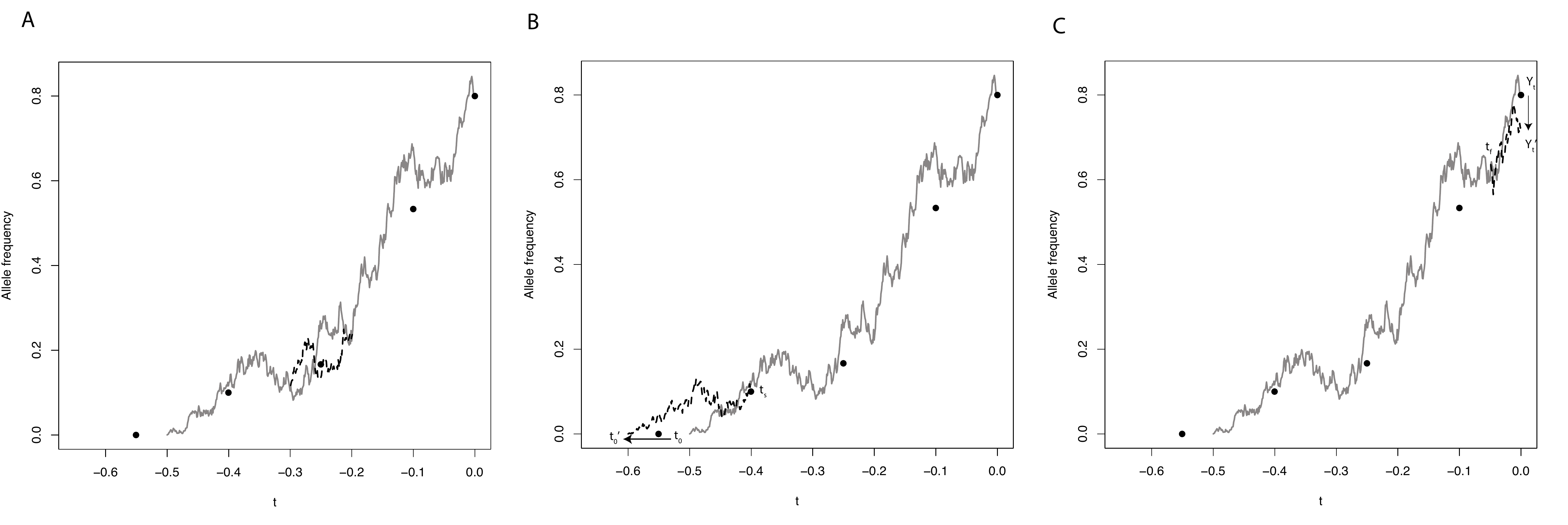} 
   \caption{Illustration of path updates. 
	Filled circles correspond to the same sample frequencies as in Figure 1. 
	The solid gray line in each panel is the current allele frequency trajectory 
	and the dashed black lines are the proposed updates. 
	In panel a, an interior section of path is proposed between points $s_1$ and $s_2$. 
	In panel b, a new allele age, $t_0'$ is proposed and a new path is drawn between $t_0'$ and $t_s$. 
	In panel c, a new most recent allele frequency $Y_{t_k}'$ is proposed and a new path is drawn between $t_f$ and $t_k$. }
\end{figure}

\subsubsection{Interior path updates}
To update a section of the allele frequency, we first choose a time $s_1 \in (t_0, t_k)$ uniformly at random, and then choose a time $s_2$ that is a fixed fraction of the path length subsequent to $s_1$. We prefer this approach of updating a fixed fraction of the path
to an alternative strategy of holding $s_2-s_1$ constant because paths for very strong selection may be quite short.
Recalling the definition of $f$ from \eqref{f_def}, 
we subsequently propose a new segment of transformed path between
the times $f(s_1)$ and $f(s_2)$ while keeping the values 
$Y_{f(s_1)}$ and $Y_{f(s_2)}$ fixed (Figure 2a). Such a path that is conditioned
to take specified values at both end-points of the interval over which it is defined is called a bridge, 
and by updating small portions of the path instead of the whole path at once, 
we are able to obtain the desirable behavior that our Metropolis-Hastings algorithm is able to stay in regions of path space with high posterior probability. If we instead drew the whole path each time, we would much less efficiently target the posterior distribution. 

Noting that bridges must be sampled against the \emph{transformed} time scale, the best bridges for the allele frequency path would be realizations of Wright-Fisher bridges themselves. However, sampling Wright-Fisher bridges is challenging (but see \cite{schraiber2013analysis,jenkins2015exact}), so we instead opt to sample bridges for the transformed path from the Bessel(0) process. Sampling Bessel(0) bridges can be accomplished by first sampling Bessel(4) bridges (as described in \cite{schraiber2013analysis}) and then recognizing that a Bessel(4) process is the same as a Bessel(0) process conditioned to never hit $0$ and hence has the same bridges  -- in the language of the general theory of Markov processes, the Bessel(0) and Bessel(4)
processes are Doob $h$-transforms of each other and it is well-known that processes related in this
way share the same bridges. 
We denote by $(Y'_\tau)_{\tau \ge 0}$ the path that has the proposed bridge spliced in between 
times $f(s_1)$ and $f(s_2)$ and coincides with $(Y_\tau)_{\tau \ge 0}$ outside the interval $[f(s_1), f(s_2)]$.

In the Appendix, we show that the acceptance probability in this case is simply
\begin{equation}
\label{interior_acceptance_probability}
\min\left\{1, \frac{L(D, (Y'_\tau)_{f(s_1)\leq \tau \leq f(s_2)} \, | \, \alpha_1, \alpha_2, t_0)}{L(D, (Y_\tau)_{f(s_1)\leq \tau \leq f(s_2)} \, | \,\alpha_1, \alpha_2, t_0)}\right\}.
\end{equation}
Note that we only need to compute the likelihood ratio for the segment of transformed
path that changed between the times $f(s_1)$ and $f(s_2)$.

\subsubsection{Allele age updates}
The first sample time with a non-zero count of the derived allele (Figure 2b)
is $t_s$, where $s = \min\{ i : c_i > 0\}$.  We must have $t_0 < t_s$.  Along with
proposing a new value $t_0'$ of the allele age 
$t_0$ we will propose a new segment of the allele frequency
path from time $t_0'$ to time $t_s$.  
Changing the allele age $t_0$ to some new proposed value $t_0'$ changes the
definition of the function $f$ in \eqref{f_def}.  Write
$f'(t) = \int_{t_0'}^t \frac{1}{\rho(s)} \, ds$, where we stress that the prime does not denote a derivative.
The proposed transformed path $Y'$ consists of a new piece of path that goes from location $0$
at time $0$ to location $Y_{f(t_s)}$ at time $f'(t_s)$ and then has $Y_{f'(t)}' = Y_{f(t)}$ for $t \ge t_s$.
We use the improper prior $\rho(t_0)$ for $t_0$, which reflects the fact that an allele is more
likely to arise during times of large population size \cite{slatkin2001simulating}.
In the Appendix, we show that the acceptance probability is
\begin{equation}
\label{allele_age_acceptance_probability}
\min\left\{1,  \frac{L(D, (Y'_\tau)_{0 \leq \tau \leq f'(t_s)} \, | \, \alpha_1, \alpha_2, t_0')}{L(D, (Y_\tau)_{0 \leq \tau \leq f(t_s)} \, | \,\alpha_1, \alpha_2, t_0)}\frac{\psi(Y_{f'(t_s)}'; f'(t_s))}{\psi(Y_{f(t_s)}; f(t_s)))}\frac{q(t_0 | t_0')}{q(t_0' | t_0)}\frac{\rho(t_0')}{\rho(t_0)}\right\}
\end{equation}
where, in the notation of Subsection~\ref{SS:path_likelihoods},
\begin{equation}
\psi(y;\epsilon) = \frac{y^2}{\epsilon^2}\exp\left\{-\frac{y^2}{2\epsilon} \right\} 
= \frac{\mathbb{Q}^0(\bar Y_\epsilon \in dy)}{dy}
\label{besselentrance}
\end{equation}
is the density of the
so-called entrance law for the Bessel(0) process that appears in the characterization of
the $\sigma$-finite measure $\mathbb{Q}^0$ and $q(t_0' | t_0)$ is the proposal distribution
of $t_0'$ (in practice, we use a half-truncated normal distribution centered at $t_0$, with the 
upper truncation occurring at the first time of non-zero observed allele frequency). 

\subsubsection{Most recent allele frequency update}
While the allele frequency at sample times $t_1, t_2, \ldots, t_{k-1}$ 
are updated implicitly by the interior path update, 
we update the allele frequency at the most recent
sample time $t_k$ separately (note that the most recent allele frequency is not an additional parameter, 
but simply a random variable with a distribution implied by the Wright-Fisher model on paths). 
We do this by first proposing a new allele frequency $Y_{f(t_k)}'$ and then proposing a new bridge from $Y_{f(t_f)}$ to $Y_{f(t_k)}'$ where $t_f \in (t_{k-1}, t_k)$ is a fixed time (Figure 2c). 
If $q(Y_{f(t_k)}' \, | \,Y_{f(t_k)})$ is the proposal density for $Y_{f(t_k)}'$ given $Y_{f(t_k)}$ (in practice, we use
a truncated normal distribution centered at $Y_{f(t_k)}$ and truncated at $0$ and $\pi$), then, arguing along the same lines as the interior path update and the allele age update, 
we accept this update with probability
\begin{equation}
\label{most_recent_allele_frequency_acceptance_probability}
\min\left\{1,  \frac{L(D, (Y'_\tau)_{f(t_f) \leq \tau \leq f(t_k)} \, | \, \alpha_1, \alpha_2, t_0)}{L(D, (Y_\tau)_{f(t_f) \leq \tau \leq f(t_k)} \, | \,\alpha_1, \alpha_2, t_0)}\frac{q(Y_{f(t_k)}|Y_{f(t_k)}')}{q(Y_{f(t_k)}'|Y_{f(t_k)})}\frac{Q(Y_{f(t_f)},Y_{f(t_k)}; f(t_k) - f(t_f))}{Q(Y_{f(t_f)},Y_{f(t_k)}'; f(t_k) - f(t_f))}\right\},
\end{equation}
where 
\begin{equation}
Q(x,y;t) = \frac{y}{t}\exp\left\{-\frac{x^2+y^2}{2t} \right\}I_1\left(\frac{xy}{t} \right)
\label{besseltransition}
\end{equation}
is the transition density of the Bessel(0) process 
(with $I_1(\cdot)$ being the Bessel function of the first kind with index $1$) -- see \cite[Section~4.3.6]{MR613983}.
Again, it is only necessary to compute the likelihood ratio for the segment of transformed path
that changed between the times $f(t_f)$ and $f(t_k)$.

\subsection{Updates to $\alpha_1$ and $\alpha_2$}
Updates to $\alpha_1$ and $\alpha_2$ are conventional scalar parameter updates. For example,
letting $q(\alpha_1' \, | \, \alpha_1)$ be the proposal density for the new value of $\alpha_1$, 
we accept the new proposal with probability
\[
\min\left\{1,  \frac{L(D, (Y_\tau)_{\tau \geq 0} \, | \, \alpha_1', \alpha_2, t_0)}
{L(D, (Y_\tau)_{\tau \geq 0} \, | \,\alpha_1, \alpha_2 , t_0)}
\frac{q(\alpha_1 \, | \, \alpha_1')}
{q(\alpha_1' \, | \,\alpha_1)}
\frac{\pi(\alpha_1', \alpha_2, t_0)}{\pi(\alpha_1, \alpha_2, t_0)}\right\}.
\]
The acceptance probability for $\alpha_2$ is similar.
For both $\alpha_1$ and $\alpha_2$, we use a heavy-tailed Cauchy prior with median 0 and scale parameter 100,
and we take the parameters $\alpha_1, \alpha_2, t_0$ to be independent under the prior distribution. In addition,
we use a normal proposal distribution, centered around the current value of the parameter. 
Here, it is necessary to compute the likelihood across the whole path. 

\section{Results}
We first test our method using simulated data to assess its performance and then apply it to two real datasets from horses. 

\subsection{Simulation performance}
To test the accuracy of our MCMC approach, we simulated allele frequency trajectories with ages uniformly distributed between $0.1$ and $0.3$ diffusion time units ago, evolving with $\alpha_1$ and $\alpha_2$ uniformly distributed between $0$ and $100$. We simulate allele frequency trajectories using an Euler approximation to the Wright-Fisher SDE \eqref{WF_SDE} with $\rho(t) \equiv 1$. At each time point between $-0.4$ and $0.0$ in steps of $0.05$, we simulated of $20$ chromosomes.

We then ran the MCMC algorithm for $1,000,000$ generations, sampling every $1000$ generations to obtain $1000$ MCMC samples for each simulation. After discarding the first 500 samples from each MCMC run as burn-in, we computed the effective sample size of the allele age estimate using the R package \texttt{coda} \cite{coda}. For the analysis of the simulations, we only included simulations that had an effective sample size greater than 150 for the allele age, resulting in retaining 744 out of 1000 simulations. 

\begin{figure}[h] 
   \includegraphics[width=.9\textwidth]{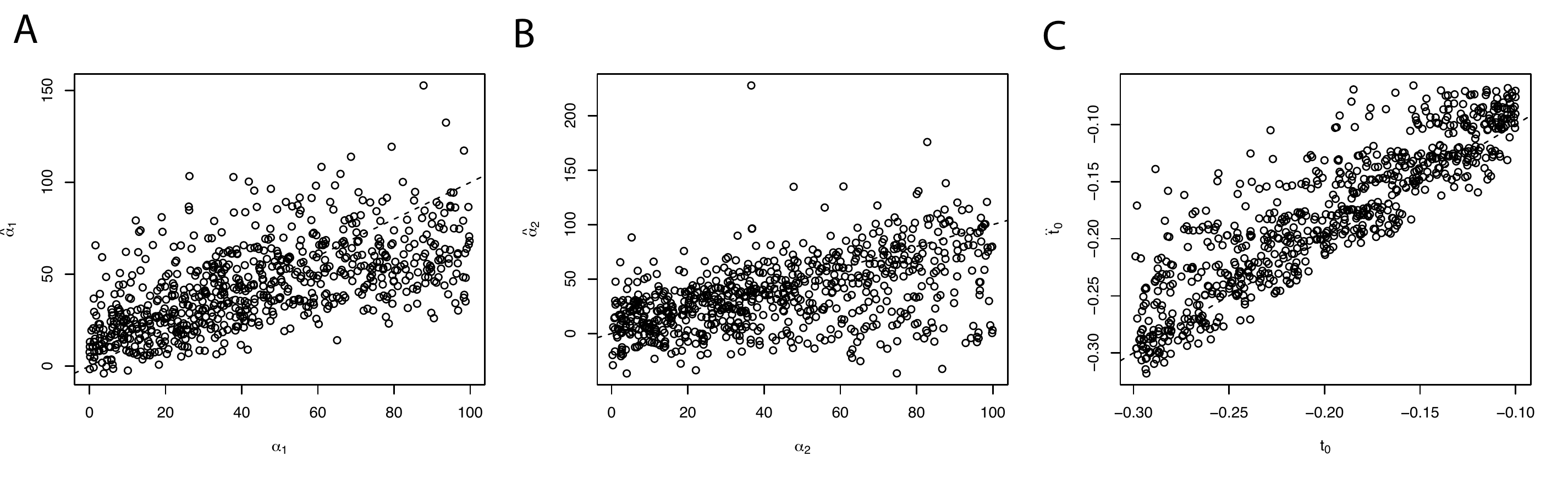} 
   \caption{Maximum \emph{a posterior} estimates of different parameters. Each panel shows the true value of a parameter on the $x$-axis, while the inferred value is on the $y$-axis. Dashed line is $y = x$.}
   \label{accuracy}
\end{figure}

Because our MCMC analysis provides a full posterior distribution on parameter values, we summarized the results by computing the maximum \emph{a posteriori} estimate of each parameter. We find that across the range of simulated $\alpha_1$ values, estimation is quite accurate (Figure \ref{accuracy}A). There is some downward bias for large true values of $\alpha_1$, indicating the influence of the prior. On the other hand, the strength of selection in favor of the homozygote, $\alpha_2$, is less well estimated, with a more pronounced downward bias (Figure \ref{accuracy}B). This is largely because most simulated alleles do not reach sufficiently high frequency for homozygotes to be common. Hence, there is very little information regarding the fitness of the homozygote. Allele age is estimated accurately, although there is a slight bias toward estimating a more recent age than the truth (Figure \ref{accuracy}C). 

\subsection{Application to ancient DNA}
 We applied our approach to real data by reanalyzing the MC1R and ASIP data from \cite{ludwig2009coat}. In contrast to earlier analyses of these data, we explicitly incorporated the demography of the domesticated horse, as inferred by \cite{sarkissian2015evolutionary}, using a generation time of $8$ years. Table \ref{horse_table} shows the sample configurations and sampling times corresponding to each locus, where diffusion units are scaled to $2N_0$, with $N_0 = 16000$ being the most recent effective size reported by 
\cite{sarkissian2015evolutionary}. For comparison, we also analyzed the data assuming the population size has been constant at $N_0$.
 
\begin{table}[h]
\begin{tabular} { c | c | c | c | c | c | c |  }
Sample time (years BCE) & 20,000 & 13,100 & 3,700 & 2,800 & 1,100 & 500 \\ \hline
Sample time (diffusion units) & 0.078 & 0.051 & 0.014 & 0.011 & 0.004 & 0.002 \\ \hline
Sample size & 10 & 22 & 20 & 20 & 36 & 38 \\ \hline
Count of ASIP alleles & 0 & 1 & 15 & 12 & 15 & 18 \\ \hline
Count of MC1R alleles & 0 & 0 & 1 & 6 & 13 & 24 
\end{tabular}
\caption{Sample information for horse data. Diffusion time units are calculated assuming $N_0 = 2500$ and a generation time of $5$ years.}
\label{horse_table}
\end{table}

\begin{figure}[h] 
   \includegraphics[width=\textwidth]{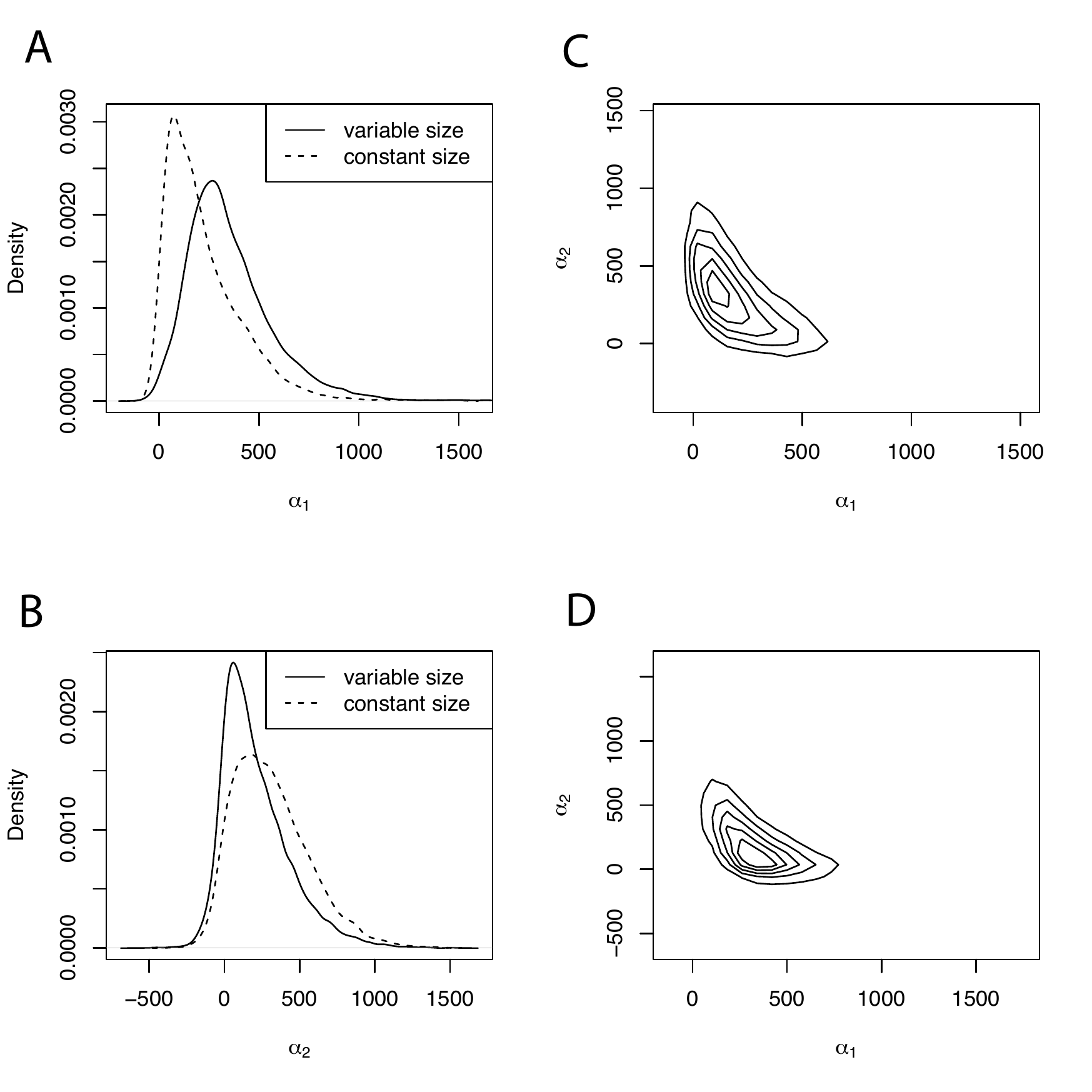} 
   \caption{Posterior distributions of selection coefficients for the MC1R locus. Panels A and B show marginal distributions of $\alpha_1$ and $\alpha_2$, respectively, with the solid line indicating the posterior obtained from an analysis including the full demographic history, and the dotted line showing what would be inferred in a constant size population. Panels C and D show contour plots of the joint distribution of $\alpha_1$ and $\alpha_2$ without and with demography, respectively.}
   \label{mc1r_alpha}
\end{figure}

With the MC1R locus, we found that posterior inferences about selection coefficients can be strongly influenced by whether or not demographic information is included in the analysis (Figure \ref{mc1r_alpha}). Marginally, we see that incorporating demographic information results in an inference that $\alpha_1$ is larger than the constant-size model (MAP estimates of 267.6 and 74.1, with and without demography, respectively; Figure \ref{mc1r_alpha}A), while $\alpha_2$ is inferred to be smaller (MAP estimates of 59.1 and 176.2, with and without demography, respectively; Figure \ref{mc1r_alpha}B). This has very interesting implications for the mode of selection inferred on the MC1R locus. With constant demography, the trajectory of the allele is estimated to be shaped by positive selection (joint MAP, $\alpha_1 = 87.6$, $\alpha_2 = 394.8$; Figure \ref{mc1r_alpha}C), while when demographic information is included, selection is inferred to act in an overdominant fashion (joint MAP, $\alpha_1 = 262.5$, $\alpha_2 = 128.1$; Figure \ref{mc1r_alpha}D). 

\begin{figure}[h] 
   \includegraphics[width=\textwidth]{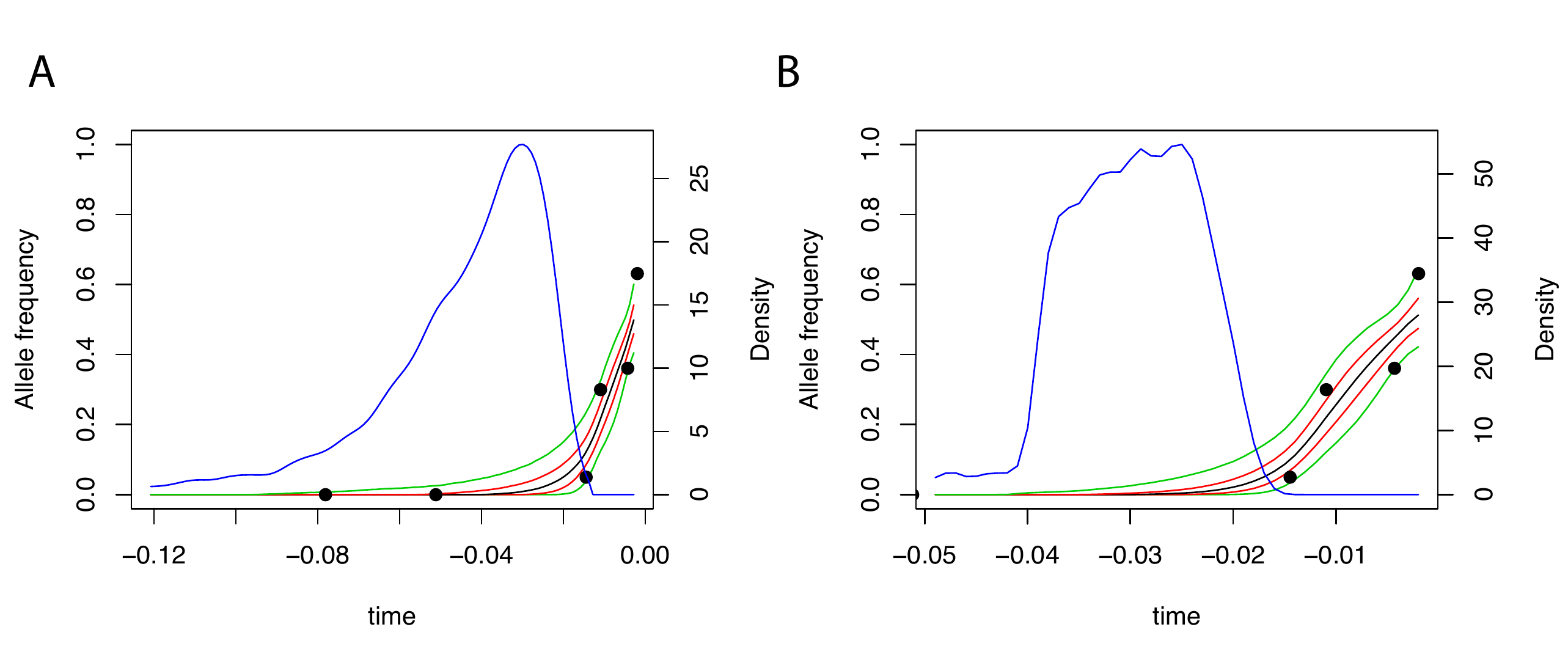} 
   \caption{Posterior distribution on allele frequency paths for the MC1R locus. Each panel shows the sampled allele frequency data (filled circles), the point-wise median (black), 25 and 75\% quantiles (red), and 5 and 95\% quantiles (green) of the posterior distribution on paths, and the posterior distribution on allele age (blue). Panel A reports inference with constant demography, while panel B shows the result of inference with the full demographic history.}
   \label{mc1r_path}
\end{figure}

Incorporation of demographic history also has substantial impacts on the inferred distribution of allele ages (Figure \ref{mc1r_path}). Most notably, the distribution of the allele age for MC1R is significantly truncated when demography is incorporated, in a way that correlates to the demographic events (Supplementary Figure \ref{mc1r_age_pop}). While both the constant-size history and the more complicated history result in a posterior mode at approximately the same value of the allele age, the domestication bottleneck inferred by \cite{sarkissian2015evolutionary} makes it far less likely that the allele rose more anciently than the recent population expansion. Because the allele is inferred to be younger under the model incorporating demography, the strength of selection in favor of the homozygote must be higher to allow it to escape low frequency quickly and reach the observed allele frequencies. Hence, $\alpha_1$ is inferred to be much higher when demographic history is explicitly modeled. 

\begin{figure}[h] 
   \includegraphics[width=\textwidth]{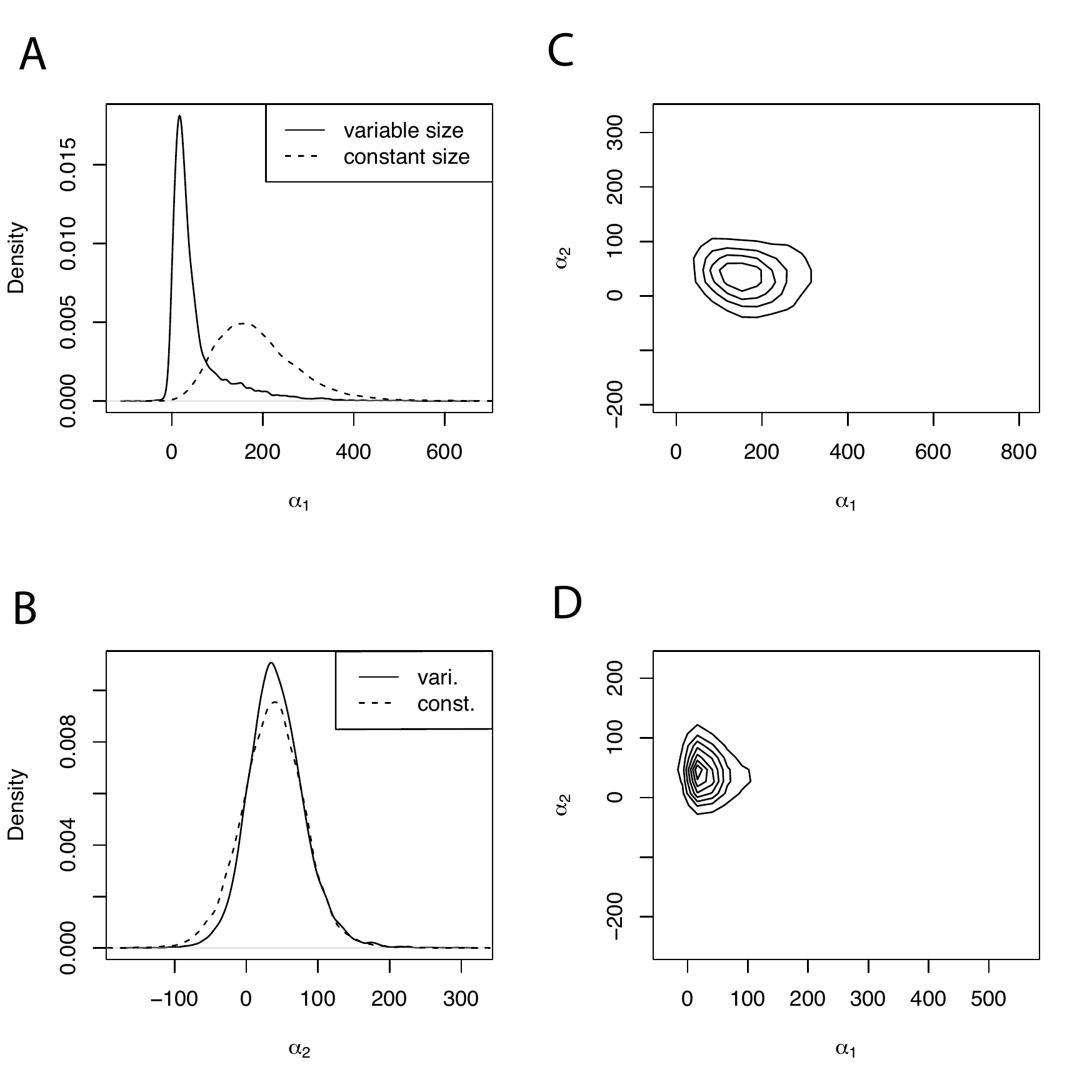} 
   \caption{Posterior distributions of selection coefficients for the ASIP locus. Panels as in Figure \ref{mc1r_alpha}}
   \label{asip_alpha}
\end{figure}

Incorporation of demographic history has an even more significant impact on inferences made about the ASIP locus (Figure \ref{asip_alpha}). Most strikingly, while $\alpha_1$ is inferred to be very large without demography, it is inferred to be close to 0 when demography is incorporated (MAP estimates of 16.3 and 159.9 with and without demography, respectively; Figure \ref{asip_alpha}A). On the other hand, inference of $\alpha_2$ is largely unaffected (MAP estimates of 34.7 and 39.8 with and without demography, respectively; Figure \ref{asip_alpha}B). Interestingly, this has an opposite implication for the mode of selection compared to the results for the MC1R locus. With a constant-size demographic history, the allele is inferred to have evolved under overdominance (joint MAP, $\alpha_1 = 153.3$, $\alpha_2  = 47$; Figure \ref{asip_alpha}C), but when the more complicated demography is modeled, the allele frequency trajectory is inferred to be shaped by positive, nearly additive, selection (joint MAP, $\alpha_1 = 16.4$, $\alpha_2 = 46.8$; Figure \ref{asip_alpha}D). 

\begin{figure}[h] 
   \includegraphics[width=\textwidth]{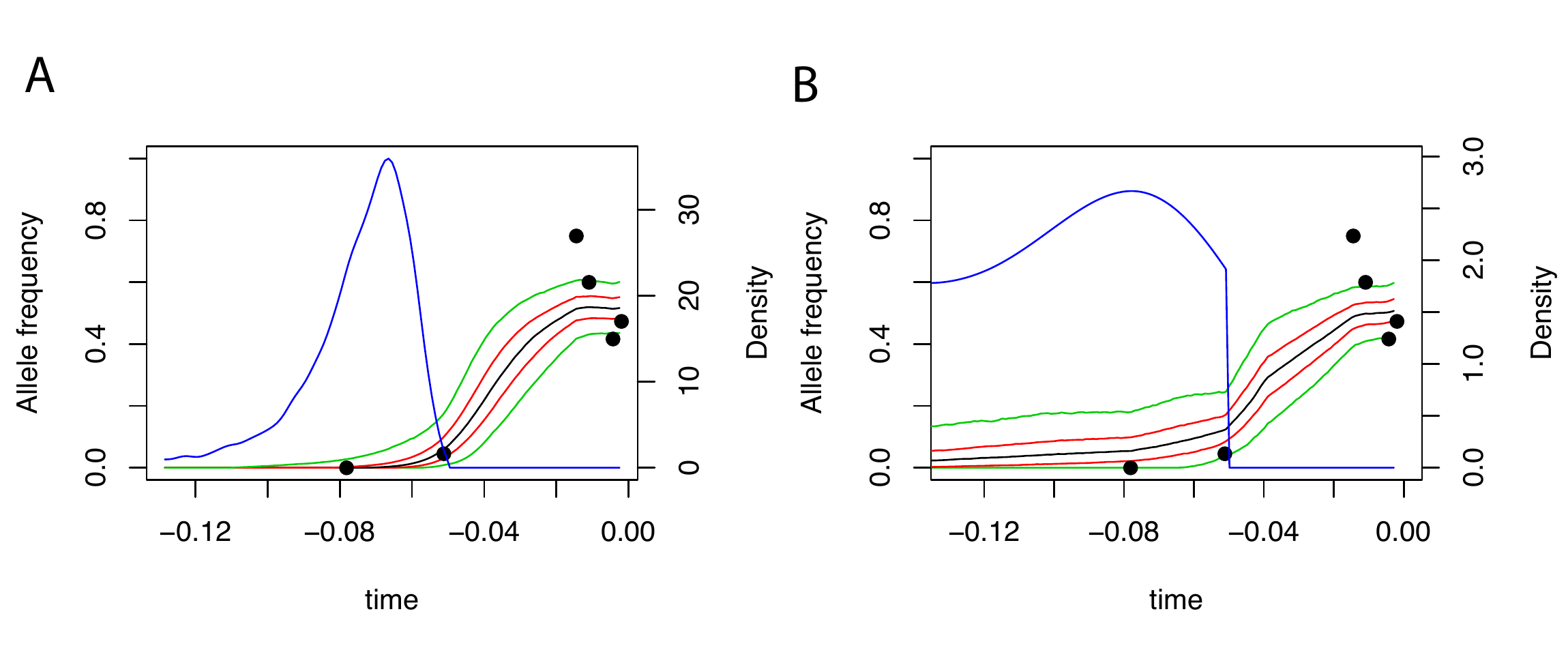} 
   \caption{Posterior distribution on allele frequency paths for the ASIP locus. Panels are as in Figure \ref{mc1r_path}.}
   \label{asip_path}
\end{figure}

Incorporating demography has a similarly opposite effect on inference of allele age (Figure \ref{asip_path}). In particular, the allele is inferred to be much older when demography is modeled, and features a multi-modal posterior distribution on allele age, with each mode corresponding to a period of historically larger population size (Figure \ref{asip_age_pop}). Because the allele is inferred to be substantially older when demography is modeled, selection in favor of the heterozygote must have been weaker than would be inferred with the much younger age. Hence, the mode of selection switches from one of overdominance in a constant demography to one in which the homozygote is more fit than the heterozygote. 

\section{Discussion}
Using DNA from ancient specimens, we have obtained a number of insights into evolutionary processes that were previously inaccessible. One of the most interesting aspects of ancient DNA is that it can provide a \emph{temporal} component to evolution that has long been impossible to study. In particular, instead of making inferences about the allele frequencies, we can directly measure these quantities. To take advantage of this new data, we developed a novel Bayesian method for inferring the intensity and direction of natural selection from allele frequency time series. In order to circumvent the difficulties inherent in calculating the transition probabilities under the standard Wright-Fisher process of selection and drift, we used a data augmentation approach in which we learn the posterior distribution on allele frequency paths. Doing this not only allows us to efficiently calculate likelihoods, but provides an unprecedented glimpse at the historical allele frequency dynamics.

The key innovation of our method is to apply high-frequency path augmentation methods \cite{roberts2001inference} to analyze genetic time series. The logic of the method is similar to the logic of a path integral, in which we average over all possible allele frequency trajectories that are consistent with the data \cite{schraiber2014path}.  
By choosing a suitable reference probability distribution against which to compute likelihood ratios, we were able to adapt these methods to infer the age of alleles and properly account for variable population sizes through time. Moreover, because of the computational advantages of the path augmentation approach, we were able to infer a model of general diploid selection. To our knowledge, ours is the first work that can estimate both allele age and general diploid selection while accounting for demography. 

Using simulations, we showed that our method performs well for strong selection and densely sampled time series. However, it is worth considering the work of \cite{watterson1979estimating}, who showed that even knowledge of the full trajectory results in very flat likelihood surfaces when selection is not strong. This is because for weak selection, the trajectory is extremely stochastic and it is difficult to disentangle the effects of drift and selection \cite{schraiber2013analysis}.

We then applied our method to a classic dataset from horses. We found that our inference of both the strength and mode of natural selection depended strongly on whether or not we incorporated demography. For the MC1R locus, a constant-size demographic model results in an inference of positive selection, while the more complicated demographic model inferred by \cite{sarkissian2015evolutionary} causes the inference to tilt toward overdominance, as well as a much younger allele age. In contrast, the ASIP locus is inferred to be overdominant under a constant-size demography, but the complicated demographic history results in an inference of positive selection, and a much older allele age. 

These results stand in contrast to those of \cite{steinrucken2014novel}, who found that the most likely mode of evolution for both loci under a constant demographic history is one of overdominance. There are a several reasons for this discrepancy. First, we computed the diffusion time units differently, using $N_0 = 16000$ and a generation time of $8$ years, as inferred by \cite{sarkissian2015evolutionary}, while \cite{steinrucken2014novel} used $N_0 = 2500$ (consistent with the bottleneck size found by \cite{sarkissian2015evolutionary}) and a generation time of 5 years. Hence, our constant-size model has far less genetic drift than the constant-size model assumed by \cite{steinrucken2014novel}. This emphasizes the importance of inferring appropriate demographic scaling parameters, even when a constant population size is assumed.  Secondly, we use MCMC to integrate over the distribution of allele ages, which can have a very long tail going into the past, while \cite{steinrucken2014novel} assume a fixed allele age.

One key limitation of this method is that it assumes that the aDNA samples all come from the same, continuous population. If there is in fact a discontinuity in the populations from which alleles have been sampled, this could cause rapid allele frequency change and create spurious signals of natural selection. Several methods have been devised to test this hypothesis \cite{sjodin2014assessing}, and one possibility would be to apply these methods to putatively neutral loci sampled from the same individuals, thus determining which samples form a continuous population. Alternatively, if our method is applied to a number of loci throughout the genome and an extremely large portion of the genome is determined to be evolving under selection, this could be evidence for model misspecification and suggest that the samples do not come from a continuous population.

An advantage of the method that we introduced is that it may be possible to extend it to incorporate information from linked neutral diversity. In general, computing the likelihood of neutral diversity linked to a selected site is difficult and many have used Monte Carlo simulation and importance sampling \cite{slatkin2001simulating, coop2004ancestral, chen2013inferring}. These approaches average over allele frequency trajectories in much same way as our method; however, each trajectory is drawn completely independently of the previous trajectories. Using a Markov chain Monte Carlo approach, as we do here, has the potential to ensure that only trajectories with a high posterior probability are explored and hence greatly increase the efficiency of such approaches.

\section{Appendix}
\subsection{A proper posterior in the limit as the intiial allele frequency approaches 0}
For reasons that we explain in Subsection~\ref{SS:path_likelihoods}, we re-parametrize
our model by replacing the path variable $(X_t)_{t \ge t_0}$ with a deterministic
time and space transformation of it $(Y_t)_{t \ge 0}$ that takes values in the interval
$[0,\pi]$ with the boundary point $0$ (resp. $\pi$) for $(Y_t)_{t \ge 0}$
corresponding to the boundary point $0$ (resp. $1$) for $(X_t)_{t \ge t_0}$.  The transformation
producing $(Y_t)_{t \ge 0}$ is such that $(X_t)_{t \ge t_0}$ can be recovered from
$(Y_t)_{t \ge 0}$ and $t_0$.

Implicit in our set-up is the initial frequency $x_0$ at time $t_0$ which corresponds
to an initial value $y_0$  at time $0$ of the transformed process $(Y_t)_{t \ge 0}$.  For the moment,
let us make the dependence on $y_0$ explicit by including it in relevant notation as a superscript.
For example, $\mathbb{P}^{y_0}( \cdot \, | \, \alpha_1, \alpha_2, t_0)$ is the prior distribution of
$(Y_t)_{t \ge 0}$ given the specified values of the other parameters $\alpha_1, \alpha_2, t_0$.
We will construct a tractable ``reference'' process $(\bar Y_t)_{t \ge 0}$ 
with distribution $\mathbb{Q}^{y_0}(\cdot)$  such that
the probability distribution $\mathbb{P}^{y_0}(\cdot \, | \,\alpha_1, \alpha_2, t_0)$ has a density 
with respect to  $\mathbb{Q}^{y_0}(\cdot)$ -- explicitly, $\mathbb{Q}^{y_0}(\cdot)$ is the
distribution of a Bessel(0) process started at location $y_0$ at time $0$.  That is,
there is a function $\Phi^{y_0}(\cdot;  \alpha_1, \alpha_2, t_0)$ on path space such that
\begin{equation}
\label{RN_deriv_positive}
\mathbb{P}^{y_0}( dy \, | \,\alpha_1, \alpha_2, t_0) 
= 
\Phi^{y_0}(y; \alpha_1, \alpha_2, t_0) \, \mathbb{Q}^{y_0}(d y)
\end{equation}
for a path $(y_t)_{t \ge 0}$.
Assuming that $\pi$ has a density with respect to Lebesgue measure which, with a slight
abuse of notation, we also denote
by $\pi$, the outcome of our Bayesian inferential procedure is determined by the ratios 
\begin{equation}
\label{likelihood_ratio}
\frac
{\mathbb{P}(dD \, | \,y^{**}, t_0^{**}) \Phi^{y_0}(y^{**}; \alpha_1^{**}, \alpha_2^{**}, t_0^{**}) 
\pi(\alpha_1^{**}, \alpha_2^{**}, t_0^{**})}
{\mathbb{P}(dD \, | \,y^*,t_0^*) \Phi^{y_0}(y^*; \alpha_1^*, \alpha_2^*, t_0^*) 
\pi(\alpha_1^*, \alpha_2^*, t_0^*)}
\end{equation}
for pairs of augmented parameter values 
$(y^*, \alpha_1^*, \alpha_2^*, t_0^*)$ and $(y^{**}, \alpha_1^{**}, \alpha_2^{**}, t_0^{**})$ 
(\emph{i.e.} the Metropolis-Hastings ratio).

Under the probability measure
$\mathbb{P}^{y_0}(\cdot \, | \,\alpha_1, \alpha_2, t_0)$, the process $(Y_t)_{t \ge 0}$
converges in distribution as $y_0 \downarrow 0$ (equivalently, $x_0 \downarrow 0$)
to the trivial process that starts at
location $0$ at time $0$ and stays there.  However, for all $\epsilon > 0$ 
the conditional distribution of $(Y_t)_{t \ge \epsilon}$ under the
probability measure $\mathbb{P}^{y_0}(\cdot \, | \,\alpha_1, \alpha_2, t_0)$ 
given the event $\{Y_\epsilon > 0\}$
converges to a non-trivial probability measure as $y_0 \downarrow 0$.
Similarly, the conditional distribution of the reference diffusion process
$(\bar Y_t)_{t \ge \epsilon}$ under the
probability measure $\mathbb{Q}^{y_0}(\cdot)$ 
given the event $\{\bar Y_\epsilon > 0\}$ converges
as $y_0 \downarrow 0$ to a non-trivial limit.  There are  $\sigma$-finite measures
$\mathbb{P}^0(\cdot \, | \,\alpha_1, \alpha_2, t_0)$ and $\mathbb{Q}^0(\cdot)$
on path space that both have infinite total mass, are such that for any $\epsilon > 0$ both of these
measures assign finite, non-zero mass to the set of paths that are strictly positive at the 
time $\epsilon$, and the corresponding conditional probability measures
are the limits as $y_0 \downarrow 0$ of the conditional probability measures described above.
Moreover, there is a function 
$\Phi^0(\cdot;  \alpha_1, \alpha_2, t_0)$ on path space such that
\begin{equation}
\label{RN_deriv_zero}
\mathbb{P}^0( dy \, | \,\alpha_1, \alpha_2, t_0) 
= 
\Phi^0(y; \alpha_1, \alpha_2, t_0) \, \mathbb{Q}^0(d y).
\end{equation}

The posterior distribution \eqref{posterior} converges to
\begin{equation}
\label{posterior_improper}
\mathbb{P}^0(d\alpha_1, d\alpha_2, dt_0; dY \, | \,D) 
= 
\frac
{\mathbb{P}(dD \, | \, Y, t_0) \mathbb{P}^0(d Y \, | \,\alpha_1, \alpha_2, t_0) \pi(d\alpha_1, d\alpha_2, dt_0)}
{\int \mathbb{P}(dD \, | \, Y') \mathbb{P}^0(dY' \, | \,\alpha_1', \alpha_2', t_0') \pi(d \alpha_1', d\alpha_2', dt_0')}.
\end{equation}
Thus, the limit as $y_0 \downarrow 0$ of a Bayesian inferential procedure 
for the augmented set of parameters can be viewed
as a Bayesian inferential procedure with the improper prior 
$\mathbb{P}^0(d Y \, | \,\alpha_1, \alpha_2, t_0) \pi(d\alpha_1, d\alpha_2, dt_0)$
for the parameters $Y, \alpha_1, \alpha_2, t_0$.
In particular, the limiting Bayesian inferential procedure is determined by the ratios 
\begin{equation}
\label{likelihood_ratio_zero}
\frac
{\mathbb{P}(dD \, | \,y^{**}, t_0^{**}) \Phi^0(h^{**}; \alpha_1^{**}, \alpha_2^{**}, t_0^{**}) 
\pi(\alpha_1^{**}, \alpha_2^{**}, t_0^{**})}
{\mathbb{P}(dD \, | \,y^*,t_0^*) \Phi^0(y^*; \alpha_1^*, \alpha_2^*, t_0^*) 
\pi(\alpha_1^*, \alpha_2^*, t_0^*)}
\end{equation}
for pairs of augmented parameter values 
$(y^*, \alpha_1^*, \alpha_2^*, t_0^*)$ and $(y^{**}, \alpha_1^{**}, \alpha_2^{**}, t_0^{**})$.

\subsection{The likelihood of the data and the path}
Write $\tau_i = f(t_i)$.  Note that $\tau_0 = f(t_0) = 0$.
Using equation \eqref{girsanovOther}, the density of the distribution of the transformed allele frequency
process $(Y_t)_{0 \le s \le \tau_k}$ against the reference distribution of the Bessel(0) process 
$(\bar Y_s)_{0 \le s \le \tau_k}$ when $Y_0 = \bar Y_0 = y_0$ can be written
\begin{equation}
\exp\left\{\int_{0}^{\tau_k}\left(a(Y_r,r)-b(Y_r)\right) \,dY_r 
- \frac{1}{2}\int_{0}^{\tau_k}\left(a^2(Y_r,r)-b^2(Y_r) \right) \, dr \right\}
\label{likeIto}
\end{equation}
where 
\[
a(y,\tau) =  - \frac{1}{2} \cot(Y_\tau)
+ \frac{1}{4}\left(\rho(f^{-1}(\tau))\sin(y)(\alpha_2+(2\alpha_1-\alpha_2)\cos(y)) \right)
\]
is the infinitesimal mean of the transformed Wright-Fisher process and 
\[
b(y) = -\frac{1}{2y}
\]
is the infinitesimal mean of the Bessel(0) process. However, as shown by \cite{sermaidis2012markov}, attempting to approximate the It\^o integral in \eqref{likeIto} using a discrete representation of the path can lead to biased estimates of the posterior distribution. Instead, consider the potential functions
\begin{align*}
H_1(y,\tau) &= \int^y a(\xi,\tau) \, d\xi \\
&= -\frac{1}{8}\left(\rho(f^{-1}(\tau))\cos^2(y)(2\alpha_1-\alpha_2)+4\log(\sin(y))\right)
\end{align*}
and
\begin{align*}
H_2(y) &= \int^y b(\xi,\tau) \, d\xi \\
&= -\frac{\log(y)}{2}.
\end{align*}
If we assume that $\rho$ is continuous (not merely right continuous with left limits), then It\^o's lemma shows that we can write
\begin{align*}
\int_{0}^{\tau_k}\left(\mu_1(Y_r,r)-\mu_2(Y_r)\right) \, dY_r 
&= H_1(Y_{\tau_k},\tau_k)-H_2(Y_{\tau_k})-\left(H_1(Y_{0},0)-H_2(Y_{0}) \right) \\
&\quad - \int_{0}^{\tau_k}\left(\frac{\partial H_1}{\partial \tau}(Y_r,r) - \frac{\partial H_2}{\partial \tau}(Y_r) \right) \, dr \\
&\quad - \int_{0}^{\tau_k}\left(\frac{\partial^2 H_1}{\partial y^2}(Y_r,r) - \frac{\partial^2 H_2}{\partial y^2}(Y_r) \right) \, dr.
\end{align*}
To generalize this to the case where $\rho$ is right continuous with left limits, write
\[
\int_{0}^{\tau_k}\left(a(Y_r,r)-b(Y_r)\right) \, dY_r = I_0 + \sum_{i=m}^{K} I_i,
\]
where $m$ and $K$ are defined in the main text,
\[
I_0 = \lim_{\tau\uparrow f(d_m)} \int_{0}^{\tau}\left(a(Y_r,r)-b(Y_r)\right) \, dY_r,
\]
for $m < i < K$,
\[
I_i = \lim_{\tau\uparrow f(d_{i+1})} \int_{f(d_i)}^{\tau}\left(a(Y_r,r)-b(Y_r)\right) \, dY_r, 
\]
and 
\[
I_K = \lim_{\tau\uparrow \tau_k} \int_{f(d_K)}^{\tau}\left(a(Y_r,r)-b(Y_r)\right) \, dY_r.
\]
It\^o's lemma can then be applied to each segment in turn. Following the conversion of the It\^o integrals into ordinary Lebesgue integrals, making the substitution $s = f^{-1}(r)$ results in the path likelihood 
displayed in \eqref{wholelikelihood}. 

\subsection{Acceptance probability for an interior path update}
When we propose a new path $(y_t')_{0 \leq t \leq \tau_k}$ to update
the current path $(y_t)_{0 \leq t \leq \tau_k}$ which doesn't
hit the boundary, the new path agrees
with the existing path outside some time interval $[v_1, v_2]$, 
and has a new segment spliced in that goes from $y_{v_1}$ at time $v_1$
to $y_{v_2}$ at time $v_2$. The proposed new path segment comes from a Bessel(0) process over
the time interval $[v_1, v_2]$ that is pinned to take the values $y_{v_1}$ and $y_{v_2}$
at the end-points; that is, the proposed new piece of path is a bridge. 

The ratio that determines the probability of accepting the proposed path is
\begin{equation}
\label{acceptance_ratio}
\frac
{
P(dD \, | \, y', t_0) 
}
{
P(dD \, | \, y, t_0)
}
\times
\frac
{ 
\mathbb{P}(dy') \kappa(dy \, | \, y')
}
{
\mathbb{P}(dy) \kappa(dy' \, | \, y),
}
\end{equation}
where $P(\cdot \, | \, y', t_0)$ and $P(\cdot \, | \, y, t_0)$ give the probability of the observed
allele counts given the transformed allele frequency paths and initial time $t_0$,
$\mathbb{P}(\cdot)$ is the distribution of the transformed Wright-Fisher diffusion starting
from $y_0 > 0$ at time $0$ (that is, the distribution we have sometimes denoted by $\mathbb{P}^{y_0}$), 
the probability kernel $\kappa(\cdot \, | \, y)$
gives the distribution of the proposed path when the current path is $y$, and 
$\kappa(\cdot \, | \, y')$ is similar.  To be completely rigorous, the second term in the product in 
\eqref{acceptance_ratio} should be interpreted as the Radon-Nikodym derivative of two
probability measures on the product of path space with itself.

Consider a finite set of times $0 \equiv \tau_0 \equiv u_0 < u_1 < \ldots < u_\ell \equiv \tau_k$.
Suppose that $\{v_1, v_2\} \in \{u_0, \ldots, u_\ell\}$  
$v_1 = u_m$ and $v_2 = u_n$  for some $m < n$.  Let $(y_t)_{0 \leq t \leq \tau_k}$ and
$(y_t')_{0 \leq t \leq \tau_k}$ be two paths that coincide on 
$[0,v_1] \cup [v_2,\tau_k] = [u_0, u_m] \cup [u_n, u_\ell]$.
Write $P(x,y;s,t)$ for the transition density (with respect to Lebesgue measure) of the 
transformed Wright-Fisher diffusion from time $s$ to time $t$ and $Q(x,y;t)$ for the
transition density (with respect to Lebesgue measure) of the Bessel(0) process.
Suppose that $(\xi, \zeta)$ is a pair of random paths with
$P((\xi, \zeta) \in (dy, dy')) = \mathbb{P}(dy) \kappa(dy' \, | \, y)$.  Then, writing
$z_t = y_t = y_t'$ for $t \in [0,v_1] \cup [v_2,\tau_k] = [u_0, u_m] \cup [u_n, u_\ell]$, we have
\[
\begin{split}
& P(\xi_{u_1} \in dy_{u_1}, \ldots, \xi_{u_\ell} \in dy_{u_\ell},
\zeta_{u_1} \in dy_{u_1}', \ldots, \zeta_{u_\ell} \in dy_{u_\ell}') \\
& \quad =
P(z_{u_0}, z_{u_1}; u_0, u_1) dz_{u_1} \times \cdots \times P(z_{u_{m-1}}, z_{u_m}; u_{m-1}, u_m) dz_{u_m} \\
& \qquad \times
P(z_{u_m}, y_{u_{m+1}}; u_{m}, u_{m+1}) dy_{u_{m+1}} \times \cdots \times P(y_{u_{n-1}}, z_{u_n}; u_{n-1}, u_n) dz_{u_n} \\
& \qquad \times
P(z_{u_n}, z_{u_{n+1}}; u_{n}, u_{n+1}) dz_{u_{m+1}} \times \cdots \times P(z_{u_{\ell-1}}, z_{u_\ell}; u_{\ell-1}, u_\ell) dz_{u_\ell} \\
& \qquad \times
Q(z_{u_m}, y_{u_{m+1}}'; u_{m+1} - u_{m}) dy_{u_{m+1}} \times \cdots \times Q(y_{u_{n-1}}, z_{u_n}; u_n -  u_{n-1}) \\
& \quad \qquad \bigg/ 
Q(z_{u_m}, z_{u_n}; u_n - u_m), \\
\end{split}
\]
where the factor in the denominator arises because we are proposing \emph{bridges} and hence conditioning
on going from a fixed location at $v_1 = u_m$ to another fixed location at $v_2 = u_n$.
Thus,
\[
\begin{split}
& \frac
{
P(\xi_{u_1} \in dy_{u_1}', \ldots, \xi_{u_\ell} \in dy_{u_\ell}',
\zeta_{u_1} \in dy_{u_1}, \ldots, \zeta_{u_\ell} \in dy_{u_\ell})
}
{
P(\xi_{u_1} \in dy_{u_1}, \ldots, \xi_{u_\ell} \in dy_{u_\ell},
\zeta_{u_1} \in dy_{u_1}', \ldots, \zeta_{u_\ell} \in dy_{u_\ell}')
} \\
& \quad =
\frac
{
\prod_{j=m}^{n-1} P(y_{u_j}', y_{u_{j+1}}'; u_{j}, u_{j+1}) / Q(y_{u_j}', y_{u_{j+1}}'; u_{j+1} - u_{j})
}
{
\prod_{j=m}^{n-1} P(y_{u_j}, y_{u_{j+1}}; u_{j}, u_{j+1}) / Q(y_{u_j}, y_{u_{j+1}}; u_{j+1} - u_{j})
}. \\
\end{split}
\]
Therefore, the Radon-Nikodym derivative appearing in \eqref{acceptance_ratio} is the ratio of 
Radon-Nikodym derivatives
\[
\frac
{\frac{d \tilde{\mathbb{P}}}{d \tilde{\mathbb{Q}}} (y')}
{\frac{d \tilde{\mathbb{P}}}{d \tilde{\mathbb{Q}}} (y)},
\] 
where $\tilde{\mathbb{P}}$ (resp. $\tilde{\mathbb{Q}}$)
is the distribution of the transformed Wright-Fisher diffusion 
(resp. the Bessel(0) process)
started at location $y_{v_1} = y_{v_1}'$ at time $v_1$
and run until time $v_2$.  The formula \eqref{interior_acceptance_probability} for the
acceptance probability associated with an interior path update follows immediately.

The above argument was carried out under the assumption that the transformed
initial allele frequency $y_0$ was strictly positive and so all the measures involved
were probability measures.  However, taking $y_0 \downarrow 0$ we see that
the formula \eqref{interior_acceptance_probability} continues to hold.  Alternatively,
we could have worked directly with the measure $\mathbb{P}^0$  in place
of $\mathbb{P}^{y_0}$.  The only difference is that
we would have to replace $P(y_0, y; 0, s)$ by the density $\phi(y; 0, s)$ of an
entrance law  for $\mathbb{P}^0$.  That is, $\phi(y; 0, s)$ has the property that
\[
\lim_{y_0 \downarrow 0} \frac{P(y_0, y'; 0, s')}{P(y_0, y''; 0, s'')} = \frac{\phi(y'; 0, s')}{\phi(y''; 0, s'')}
\]
for all $y', y'' > 0$ and $s', s'' > 0$ so that
\[
\int \phi(y; 0, s) P(y, z; s, t) \, dy = \phi(z; 0, t)
\]
for $0 < s < t$.  Such a density, and hence the corresponding entrance law, is unique up to a multiplicative
constant.  In any case, it is clear that the choice of entrance law in the definition
of $\mathbb{P}^0$ does not affect the formula \eqref{interior_acceptance_probability}
as the entrance law densities ``cancels out''.

\subsection{Acceptance probability for an allele age update}
The argument justifying the formula \eqref{allele_age_acceptance_probability} for the probability of
accepting a proposed update to the allele age $t_0$ is similar to the one just given for interior
path updates.  Now, however, we have to consider replacing a path $y$ that starts from $y_0$ at time $0$
and runs until time $f(t_k)$ with a path $y'$ that starts from $y_0$ at time $0$ and runs until time
$f'(t_k)$.  Instead of removing an internal segment of path and replacing it by one of the same
length with the same values at the endpoints, we replace the initial segment of path that runs from time
$0$ to $f(t_s) = \int_{t_0}^{t_s} \frac{1}{\rho(s)} \, ds$ by one that runs from time $0$ to time 
$f'(t_s) = \int_{t_0'}^{t_s} \frac{1}{\rho(s)} \, ds$, with $y_{f'(t_s)}' = y_{f(t_s)}$.

By analogy with the previous subsection, we need to consider
\[
\frac{P(\xi \in dy', T_0^\xi \in dt', \zeta \in dy, T_0^\zeta \in dt)}{P(\xi \in dy, T_0^\xi \in dt, \zeta \in dy', T_0^\zeta \in dt')},
\]
where $\xi$ is a transformed Wright-Fisher process starting at $y_0$ at time $0$ and run
to time $F^\xi = \int_{T_0^\xi}^{t_s} \frac{1}{\rho(s)} \, ds$, where $P(T_0^\xi \in dt) = \rho(t) \, dt$,
and conditional on $\xi$, $\zeta$ is a Bessel(0) bridge run from $y_0$ at time $0$ to $\xi_{F^\xi}$ 
at time
$F^\zeta = \int_{T_0^\zeta}^{t_s} \frac{1}{\rho(s)} \, ds$, where $P(T_0^\zeta \in dt) =   \rho(t) dt$
independent of $\xi$ and $T_0^\xi$.

Suppose that $0 = u_0 < u_1 < \ldots < u_m = \int_{t'}^{t_s} \frac{1}{\rho(s)} \, ds$ and
$0 = v_0 < v_1 < \ldots < v_n = \int_{t}^{t_s} \frac{1}{\rho(s)} \, ds$. We have for
$y_0', \ldots y_m'$ and $y_0, \ldots, y_n$ with $y_0 = y_0'$ and $y_m' = y_n$ that
\[
\begin{split}
&\frac{
P(
  \xi_{u_i} \in dy_i', 1 \le i \le m-1, T_0^\xi   \in dt', 
\zeta_{v_j} \in dy_j,  1 \le j \le n,   T_0^\zeta \in dt)
}
{
P(
  \xi_{v_j} \in dy_j,  1 \le j \le n-1, T_0^\xi   \in dt, 
\zeta_{u_i} \in dy_i', 1 \le i \le m,   T_0^\zeta \in dt')
} \\
& \quad =
\Bigg \{
\prod_{i=0}^{m-1} P(y_j', y_{j+1}'; u_i, u_{i+1}) \, dy_{i+1}'
\times \rho(t') \, dt' \\
& \qquad \times
\left[\prod_{j=0}^{n-2} Q(y_j,y_{j+1}; v_{j+1}-v_j) \, dy_{j+1} 
\times Q_(y_{n-1}, y_n; v_n - v_{n-1}) 
\bigg / Q_(y_0, y_n; v_n)\right]
\times dt
\Bigg\} \\
& \qquad \quad \Bigg /
\Bigg\{
\prod_{j=0}^{n-1} P(y_j, y_{j+1}; v_j, v_{j+1}) \, dy_{j+1})
\times \rho(t) \, dt \\
& \qquad \qquad \times
\left[\prod_{i=0}^{m-2} Q(y_i',y_{i+1}'; u_{i+1}-u_i) \, dy_{i+1}' 
\times Q_(y_{m-1}', y_m'; u_m - u_{m-1})
\bigg / Q_(y_0', y_m'; u_m)\right]
\times dt'
\Bigg \} \\
& \quad =
\Bigg\{
\prod_{i=0}^{m-1} P(y_j', y_{j+1}'; u_i, u_{i+1}) \, dy_{i+1}'
\times \rho(t') \, dt' \\
& \qquad \times
\left[\prod_{j=0}^{n-1} Q(y_j,y_{j+1}; v_{j+1}-v_j) \, dy_{j+1} 
\bigg / Q_(y_0, y_n; v_n)\right]
\times dt
\Bigg \} \\
& \qquad \quad \Bigg /
\Bigg \{
\prod_{j=0}^{n-1} P(y_j, y_{j+1}; v_j, v_{j+1}) \, dy_{j+1})
\times \rho(t) \, dt \\
& \qquad \qquad \times
\left[\prod_{i=0}^{m-1} Q(y_i',y_{i+1}'; u_{i+1}-u_i) \, dy_{i+1}' 
\bigg / Q_(y_0', y_m'; u_m)\right]
\times dt'
\Bigg \} \\
& \quad =
\frac
{
\prod_{i=0}^{m-1} P(y_j', y_{j+1}'; u_i, u_{i+1}) \, dy_{i+1}'
\bigg /
\left[\prod_{i=0}^{m-1} Q(y_i',y_{i+1}'; u_{i+1}-u_i) \, dy_{i+1}' 
\right]
}
{
\prod_{j=0}^{n-1} P(y_j, y_{j+1}; v_j, v_{j+1}) \, dy_{j+1})
\bigg /
\left[\prod_{j=0}^{n-1} Q(y_j,y_{j+1}; v_{j+1}-v_j) \, dy_{j+1} 
\right]
} \\
& \qquad \times \frac{Q_(y_0', y_m'; u_m)}{Q_(y_0, y_n; v_n)}
\times \frac{\rho(t')}{\rho(t)},\\
\end{split}
\]
where the second equality follows from the fact that $y_n = y_m'$.  

Thus,
\[
\begin{split}
& \frac{P(\xi \in dy', T_0^\xi \in dt', \zeta \in dy, T_0^\zeta \in dt)}{P(\xi \in dy, T_0^\xi \in dt, \zeta \in dy', T_0^\zeta \in dt')} \\
& \quad =
\frac
{\frac{d \check{\mathbb{P}}}{d \check{\mathbb{Q}}} (y')}
{\frac{d \hat{\mathbb{P}}}{d \hat{\mathbb{Q}}} (y)}
\times \frac{Q(y_0, y_{T'}'; T')}{Q(y_0, y_{T}; T)}
\times \frac{\rho(t')}{\rho(t)},
\end{split}
\]
where $T = \int_{t}^{t_s} \frac{1}{\rho(s)} \, ds$ and $T' = \int_{t'}^{t_s} \frac{1}{\rho(s)} \, ds$,
$\hat{\mathbb{P}}$ (resp. $\check{\mathbb{P}}$) is the distribution of the transformed Wright-Fisher
diffusion starting at location $y_0$ at time $0$ and run until time $T$ (resp. $T'$), and
$\hat{\mathbb{Q}}$ (resp. $\check{\mathbb{Q}}$) is the distribution of the Bessel(0)
process starting at location $y_0$ at time $0$ and run until time $T$ (resp. $T'$).

We have thusfar assumed that $y_0$ is strictly positive.  As in the previous subsection, we can let 
$y_0 \downarrow 0$ to get an expression in terms of Radon-Nikodym derivatives of
$\sigma$-finite measures and the density $\psi(y; s)$ of an
entrance law for $\mathbb{Q}^0$.  That is, $\psi(y; s)$ has the property that
\[
\lim_{y_0 \downarrow 0} \frac{Q(y_0, y'; s')}{Q(y_0, y''; s'')} = \frac{\psi(y'; s')}{\psi(y''; s'')}
\]
for all $y', y'' > 0$ and $s', s'' > 0$, so that
\[
\int \psi(y; s) Q(y, z; t) \, dy = \psi(z; s+t)
\]
for $s,t > 0$.  Up to an irrelevant multiplicative constant, $\psi$ is
given by the expression \eqref{besselentrance}, and the
formula \eqref{allele_age_acceptance_probability} for the acceptance probability follows immediately.

\subsection{Acceptance probability for a most recent allele frequency update}
The derivation of formula \eqref{most_recent_allele_frequency_acceptance_probability} for the
probability of accepting a  proposed update to the most recent allele frequency is
similar to those for the other acceptance probabilities
\eqref{interior_acceptance_probability} and \eqref{allele_age_acceptance_probability},
so we omit the details.

\section{Supplementary Figures}

\setcounter{figure}{0}

\begin{figure}[h] 
   \includegraphics[width=.9\textwidth]{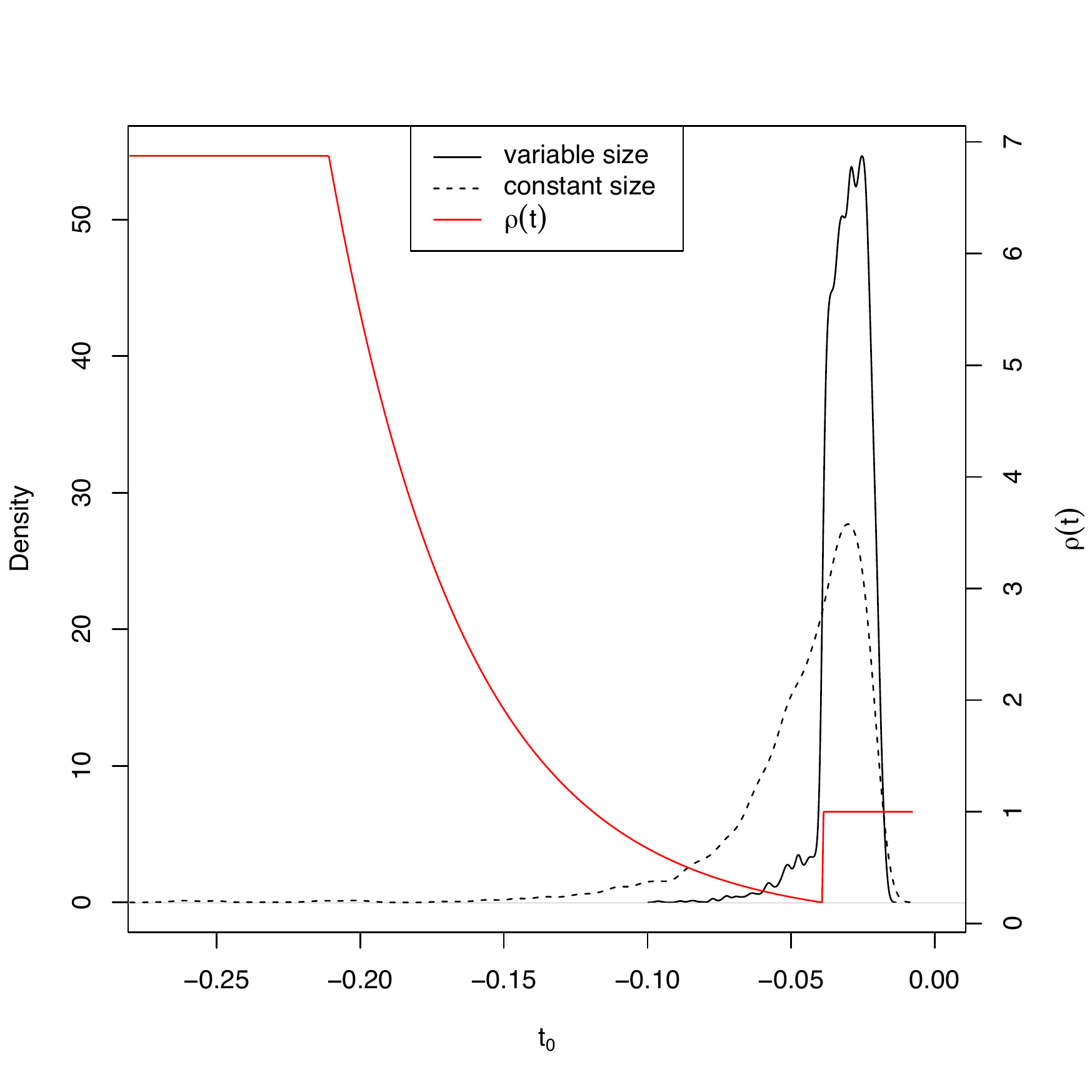} 
   \caption{Influence of population size on age estimates of the MC1R locus. The solid and dashed lines show the posterior distribution on allele age with and without demography, respectively. In red, the demographic history inferred by \cite{sarkissian2015evolutionary}.}
   \label{mc1r_age_pop}
\end{figure}

\begin{figure}[h] 
   \includegraphics[width=.9\textwidth]{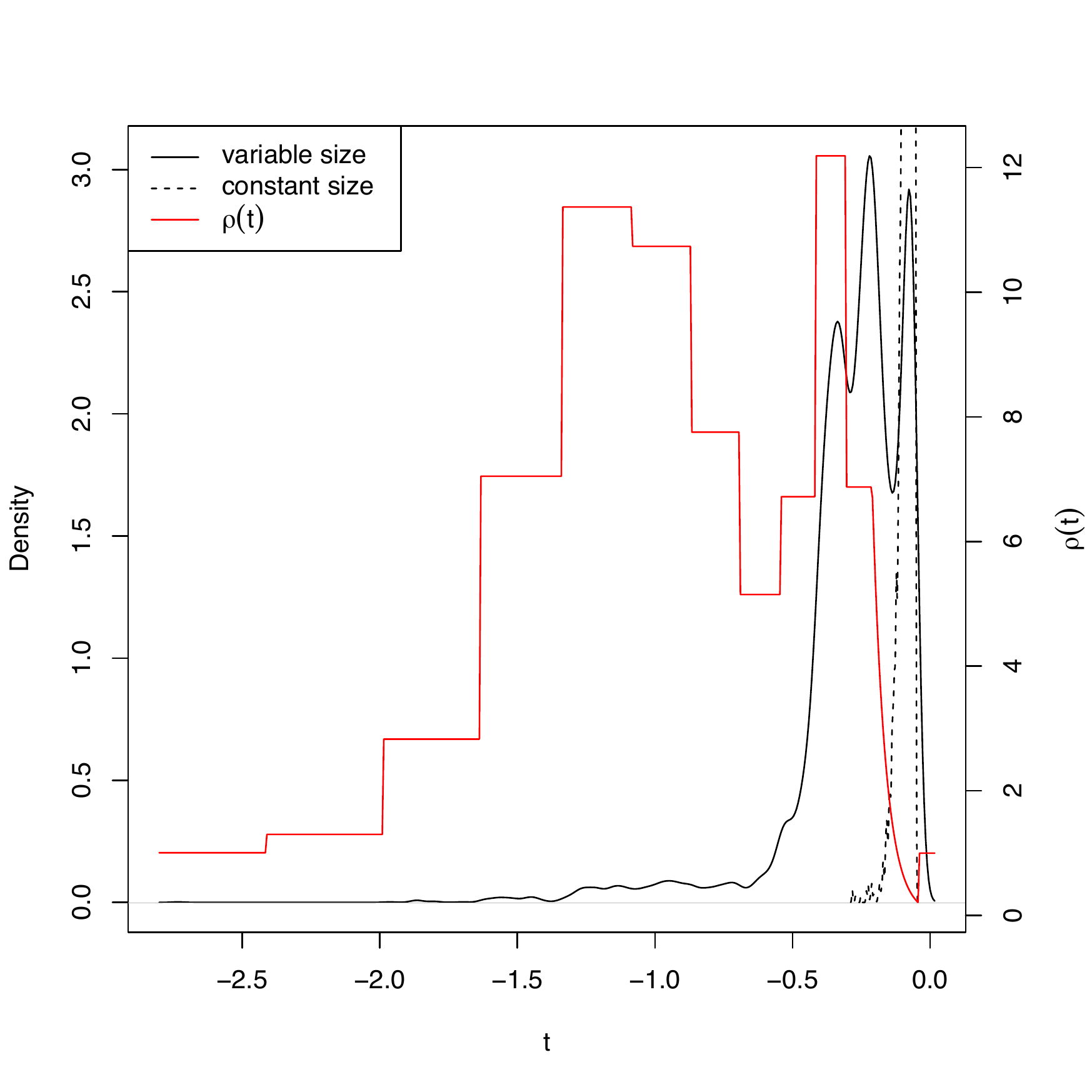} 
   \caption{Influence of population size on age estimates of the ASIP locus. Data presented is as in Figure \ref{mc1r_age_pop}}
   \label{asip_age_pop}
\end{figure}
 
 
 \newcommand{\etalchar}[1]{$^{#1}$}
\providecommand{\bysame}{\leavevmode\hbox to3em{\hrulefill}\thinspace}
\providecommand{\MR}{\relax\ifhmode\unskip\space\fi MR }
\providecommand{\MRhref}[2]{%
  \href{http://www.ams.org/mathscinet-getitem?mr=#1}{#2}
}
\providecommand{\href}[2]{#2}

\end{document}